\documentclass[11pt, a4paper]{article}
\pdfoutput=1
\usepackage{jheparxiv}
\usepackage[utf8]{inputenc}
\usepackage{amsmath,amsfonts,amssymb,graphicx}
\usepackage[normalem]{ulem}
\usepackage{mathrsfs}
\usepackage{graphicx}
\usepackage{color}
\usepackage{slashed}
\usepackage{tikz}
\usepackage{verbatim}

\newcommand{\be}{\begin{equation}}
\newcommand{\ee}{\end{equation}}
\newcommand{\bi}{\begin{itemize}}
\newcommand{\ei}{\end{itemize}}
\newcommand{\bea}{\begin{eqnarray}}
\newcommand{\eea}{\end{eqnarray}}
\newcommand{\tr}{\text{tr}\,}

\newcommand{\ud}{\mathrm{d}}
\newcommand{\LCm}{{\scriptscriptstyle -}} 
\newcommand{\LCp}{{\scriptscriptstyle +}}

\newcommand{\LCperp}{{\scriptscriptstyle \perp}}

\newcommand{\sfi}{{\sf i}}
\newcommand{\sfa}{{\sf a}}

\newcommand{\la}{\langle}
\newcommand{\ra}{\rangle}

\newcommand{\sfT}{{\mathsf T}}
\renewcommand{\d}{\mathrm{d}}
\newcommand{\im}{\mathrm{i}}
\newcommand{\cE}{\mathcal{E}}
\newcommand{\e}{\mathrm{e}}
\newcommand{\cV}{\mathcal{V}}

\title{Classical and quantum double copy of back-reaction}

\author[1]{Tim Adamo}
\emailAdd{t.adamo@ed.ac.uk}

\affiliation[1]{School of Mathematics \\
        University of Edinburgh, EH9 3FD, United Kingdom}
        
\author[2]{\& Anton Ilderton}
\emailAdd{anton.ilderton@plymouth.ac.uk}

\affiliation[2]{Centre for Mathematical Sciences \\
		University of Plymouth, PL4 8AA, United Kingdom}

\abstract{
We consider radiation emitted by colour-charged and massive particles crossing strong plane wave backgrounds in gauge theory and gravity. These backgrounds are treated exactly and non-perturbatively throughout. We compute the back-reaction on these fields from the radiation emitted by the probe particles: classically through background-coupled worldline theories, and at tree-level in the quantum theory through three-point amplitudes. Consistency of these two methods is established explicitly. We show that the gauge theory and gravity amplitudes are related by the double copy for amplitudes on plane wave backgrounds. Finally, we demonstrate that in four-dimensions these calculations can be carried out with a background-dressed version of the massive spinor-helicity formalism.
}

\begin{document}
\maketitle

\section{Introduction}\label{SECT:intro}

The \emph{double copy} is a non-trivial set of relationships between the perturbative scattering amplitudes of gauge theory and gravity, whereby the amplitudes of a gravitational theory can be expressed as the `square' of amplitudes at the same perturbative order and particle number in a gauge theory~\cite{Kawai:1985xq,Bern:2008qj,Bern:2010ue,Bern:2010yg}. When this can be made precise, double copy is an extremely powerful statement, since perturbative gravity is based on a non-polynomial Lagrangian with an infinite number of interaction vertices (e.g., the Einstein-Hilbert Lagrangian) whereas the standard perturbation theory of gauge theories is much less complicated. For scattering amplitudes on a trivial background, double copy has been formulated (see~\cite{Bern:2019prr} for a comprehensive review with references) at tree-level and beyond for a wide array of gravitational theories (the most recent addition to this catalog being massive gravity~\cite{Momeni:2020vvr,Johnson:2020pny}).

A perturbative version of double copy has also been found relating the asymptotic radiation resulting from classical dynamics of colour and gravitational sources moving in an initially flat background~\cite{Luna:2016due,Goldberger:2016iau,Luna:2016hge,Goldberger:2017frp,Luna:2017dtq,Chester:2017vcz,Goldberger:2017vcg,Goldberger:2017ogt,Shen:2018ebu,Goldberger:2019xef}. In unison with the `standard' double copy for scattering amplitudes in quantum field theory, these tools are now having a real impact on the study of gravitational wave physics. For instance, classical information at high post-Minkowski order can be systematically extracted from the relativistic amplitudes~\cite{Cheung:2018wkq}, thereby enabling new precision results for the conservative two-body Hamiltonian~\cite{Bern:2019nnu,Bern:2019crd}.



Despite these many advances, work to-date has mostly focused on perturbation theory on trivial backgrounds, but if double copy is a general property of gauge theory and gravity, then it should hold on an \emph{arbitrary} background. While there is still a limited understanding of how double copy applies to observables in general backgrounds, concrete progress has been made for specific backgrounds. It has been shown by studying three- and four-point amplitudes on plane wave backgrounds that a generalization of double copy and colour-kinematics duality does persist in the presence of background curvature\footnote{Other studies of double copy in non-trivial backgrounds can be found in~\cite{Bahjat-Abbas:2017htu,Carrillo-Gonzalez:2017iyj,Farrow:2018yni,Borsten:2019prq}.}~\cite{Adamo:2017nia,Adamo:2018mpq}. 

Both colour/electromagnetic plane waves and gravitational plane waves are highly symmetric solutions of the corresponding vacuum equations. This allows for the exact treatment of arbitrarily strong backgrounds, ensures the existence of a well-defined S-matrix, and has recently enabled calculations approaching the generality achieved in trivial backgrounds. For instance, the singularity structure of amplitudes in QED on an electromagnetic plane wave background is tightly constrained by gauge invariance~\cite{Ilderton:2020rgk}, and all-multiplicity formulae for tree-level MHV scattering of gluons and gravitons in chiral plane waves can be found~\cite{Adamo:2020syc}.

\medskip

In this paper, we consider leading-order back-reaction effects on gauge and gravitational plane wave backgrounds, and the double copy between them. The back-reaction is sourced by a colour-charged or massive scalar particle traversing the background. We compute this back-reaction both classically -- using background-coupled worldline theories -- and at tree-level in quantum field theory -- using three-point `non-linear Compton' scattering amplitudes. The classical limit of the non-linear Compton amplitudes is shown to produce the same formulae as the classical worldline calculation. We then use the prescription of~\cite{Adamo:2017nia} to show that the back-reaction corresponding to gluon emission double copies to the back-reaction corresponding to graviton emission. Finally, these results are translated into a background-dressed version of the spinor helicity formalism for massless and massive particles in four-dimensions.

This paper is organised as follows. In Section~\ref{SECT:plane-wave} we introduce and summarize the main properties of plane wave backgrounds. In Section~\ref{SECT:YM} we consider classical colour charge dynamics in a Yang-Mills plane wave background, constructing the gluon radiation field to leading order in the coupling, but \emph{exactly} in the strong background. Analogously, Section~\ref{SECT:GR} deals with the leading order classical back-reaction on a gravitational plane wave due to radiation from a massive scalar particle. In Section~\ref{NLComp} we turn to the quantum theory, constructing the tree-level three-point non-linear Compton amplitudes for gluon and graviton emission from massive scalars crossing Yang-Mills and gravitational plane waves, respectively. The classical limit of these non-linear Compton amplitudes are shown to give the same formulae as obtained in Section~\ref{SECT:YM} for classical back-reaction in gauge theory and gravity.

Section~\ref{SECT:DC} demonstrates that the non-linear Compton amplitudes obey the double copy prescription of~\cite{Adamo:2017nia} for plane wave backgrounds, and observes similarities with the perturbative double copy of~\cite{Goldberger:2016iau,Luna:2016hge} in the classical limit. In Section~\ref{SECT:SHF}, we consider the restriction of the non-linear Compton amplitudes to four space-time dimensions, where the spinor helicity formalism can be employed\footnote{Subsection~\ref{SSECT:form} can be read independently as an introduction to the spinor helicity formalism for general use in plane wave backgrounds.}. We introduce a background-dressed version of this formalism which describes on-shell kinematics for both massless and massive particles (in gauge and gravitational backgrounds), and translate our results for non-linear Compton scattering and the double copy into this formalism. Section~\ref{SECT:CONCS} concludes with a discussion of future directions.

\section{Plane wave backgrounds}\label{SECT:plane-wave}
Our aim is to study emission and back-reaction at lowest order, as massive particles cross a strong radiation background field in gauge theory and general relativity. These strong backgrounds will be treated exactly as \emph{plane waves}, which are highly-symmetric solutions of the vacuum equations of motion~\cite{Baldwin:1926,volkov35,Ehlers:1962zz,Trautman:1980bj,Ilderton:2018lsf,Zhang:2019gdm}. In $d$ space-time dimensions, plane waves possess $2d-3$ symmetries, one of which is covariantly constant and null. The generators of these symmetries form a Heisenberg algebra with centre $n$, the generator of the covariantly constant null symmetry. In both gauge theory and gravity, plane waves are characterised by functional degrees of freedom which correspond in form and number to the on-shell degrees of freedom for a gluon or graviton; as such plane waves can be viewed as coherent superpositions of the gauge bosons.

\subsubsection*{Gauge theory}

In the gauge theory context, we work in lightfront coordinates on Minkowski space
\be\label{lcMink}
\d s^2=2\,\d x^{+}\,\d x^{-}-\d x_{a}\,\d x^{a}=2\,\d x^{+}\,\d x^{-}-\d x_{\perp}\,\d x^{\perp}\,,
\ee
where the $d-2$ transverse directions $x^{a}$, $a=1,\ldots,d-2$ are often abbreviated by $x^{\perp}$. A plane wave is described (up to gauge transformations) by a gauge potential
\be\label{PGgauge1}
\mathsf{a}_{\mu}=x^{\perp}\,\dot{a}_{\perp}(x^-)\,n_{\mu}\,,
\ee
where $n_{\mu}=\delta_{\mu}^{-}$, $a_{\perp}$ are $d-2$ functions of the lightfront variable $x^-$, valued in the Cartan of the gauge group~\cite{Trautman:1980bj,Adamo:2017nia}, and $\dot{a}_{\perp}=\partial_{-}a_{\perp}$. The covariantly constant null symmetry associated with this solution is generated by the vector dual to $n_{\mu}$:
\be\label{nulliso}
n=\frac{\partial}{\partial x^{+}}\,.
\ee
It is easy to see that \eqref{PGgauge1} solves the vacuum Yang-Mills equations for any Cartan-valued~$a_{\perp}$; however, we also demand that $\dot{a}_{\perp}$ is compactly supported in $x^-$. This `sandwich' condition says that the plane wave has finite duration in lightfront time, and is physically motivated: it ensures that, as a background field, the plane wave admits a well-defined S-matrix~\cite{schwinger51,Adamo:2017nia}.

Using the explicit form of the potential \eqref{PGgauge1}, the field strength of the background is
\be\label{fmunudef}
	f_{\mu\nu} = \dot{a}_{\mu}(x^-)\,n_{\nu}-n_{\mu}\,\dot{a}_{\nu}(x^-)\,,
\ee
where $a_\mu\equiv \delta_\mu^\LCperp a_\LCperp$ throughout. 

\subsubsection*{Gravity}

A plane wave in general relativity is described by a metric with a single non-trivial component (in Brinkmann coordinates~\cite{Brinkmann:1925fr}):
\be\label{grPW1}
\d s^2=2\,\d x^{+}\,\d x^{-}-\d x_{a}\,\d x^{a}-H_{ab}(x^-)\,x^{a}x^{b}\,(\d x^-)^2\,.
\ee
It is easy to see that the null vector $n=\partial_{+}$ is a covariantly constant Killing vector of this metric. The only condition required for this metric to solve the vacuum Einstein equations is that the symmetric $(d-2)\times(d-2)$ matrix of functions $H_{ab}(x^-)$ is trace-free: $H^{a}_{a}(x^-)=0$. We also impose the sandwich condition, which here means that $H_{ab}$ is compactly supported in $x^-$, to ensure a well-defined S-matrix in the plane wave space-time~\cite{Gibbons:1975jb}.

There are various geometric structures associated with the plane wave metric \eqref{grPW1} which play an important role. The geodesic deviation equation for transverse coordinates in this space-time takes the form
\be\label{geodev}
\ddot{e}_{a}(x^-)=H_{ab}(x^-)\,e^{b}(x^-)\,, \qquad e^{a}:=\Delta x^{a}\,.
\ee
Taking a set of $d-2$ linearly-independent solutions to this geodesic equation defines a vielbein $E^{i}_{a}(x^-)$ (and inverse $E_{i\,a}$), where the index $i=1,\ldots,d-2$ runs over the independent solutions. This vielbein obeys
\be\label{vielbein}
\ddot{E}^{i}_{a}=H_{ab}\,E^{i\,b}\,, \qquad \dot{E}^{a}_{[i}\,E_{j]\,a}=0\,,
\ee
and defines a metric on the space of solutions to the geodesic deviation equation \eqref{geodev}:
\be\label{transmet}
\gamma_{ij}(x^-):=E^{a}_{(i}\,E_{j)\,a}\,.
\ee
The vielbein encodes the geometric optics associated with the null geodesic congruence generated by
\be\label{nullcong}  
\frac{\partial}{\partial x^{-}}+\frac{x^{a}x^{b}}{2}\left(H_{ab}-\dot{E}^{i}_{a}\,\dot{E}_{i\,b}\right)\frac{\partial}{\partial x^+}+\dot{E}^{a}_{i}\,E^{i}_{b} x^{b}\,\frac{\partial}{\partial x^{a}}\,,
\ee
through the deformation tensor
\be\label{deftensor}
\sigma_{ab}(x^{-}):=\dot{E}^{i}_{a}\,E_{i\,b}\,.
\ee
Decomposing $\sigma_{ab}$ into its trace and trace-free parts gives the expansion and shear of the null geodesic congruence \eqref{nullcong}, respectively.

Note that the metric \eqref{grPW1} is written in Kerr-Schild form, and as such is immediately suitable for a description in terms of the classical double copy between algebraically special solutions of gauge theory and gravity~\cite{Monteiro:2014cda}. Sure enough, the `single copy' of this metric is precisely the gauge theoretic plane wave \eqref{PGgauge1}.

\section{Classical colour charge dynamics}
\label{SECT:YM}
We consider here a single colour-charged particle coupled to classical Yang-Mills fields. As well as its position $x^\mu \equiv x^\mu(\tau)$, parameterised by proper time $\tau$, the particle carries colour degrees of freedom $c^\sfa(\tau)$ in the adjoint. The governing (coupled) equations are the Yang-Mills equations for the gauge field $A_\mu^\sfa$, the Lorentz force law for the particle orbit $x^\mu(\tau)$, and Wong's equation~\cite{Wong:1970fu} for the colour degrees of freedom, respectively\footnote{$D$ represents the usual covariant derivative in this expression only; elsewhere it stands for the covariant derivative with respect to a background gauge field.}:
\bea
\label{Maxwell}
	D^\mu F_{\mu\nu}^\sfa &=& -g\, J_{\nu}^{\sfa} \;, \qquad \text{where}\qquad J_{\nu}^{\sfa}(x) = \int\!\ud \tau\, c^\sfa(\tau) \dot{x}_\nu(\tau) \delta^d(x-x(\tau)) \;, \\
\label{Lorentz}
	m\, \ddot{x}_\mu &=& -g\, c^\sfa\, F_{\mu\nu}^\sfa\, \dot{x}^\nu \;, \\
\label{Wong}
	\dot{c}^\sfa  &=& -g\, \mathrm{f}^{\mathsf{abc}}\, {\dot x}^\mu\, A_\mu^\mathsf{b}\, c^\mathsf{c}  \;,
\eea
where $g$ is the Yang-Mills coupling constant and $\mathrm{f}^{\mathsf{abc}}$ are the structure constants of the gauge group. The physical situation of interest is that the charge interacts with a given background field $A^\mu_\text{in}\equiv\sfa^\mu$, which is initially present and obeys the vacuum Yang-Mills equations. We take this background field to be given by the sandwich plane wave from Section~\ref{SECT:plane-wave}. The particle is accelerated by the background and emits (colour) radiation: in other words there is back-reaction on $\sfa_\mu$. Comparing with~\cite{Goldberger:2016iau}, the main difference in our setup is the presence of the background, which is what allows us to generate non-trivial radiation with only a single colour charge.

We allow the background to be arbitrarily strong, meaning $g \sfa_\mu$ is characterised by a dimensionless coupling greater than one, and so cannot be treated in perturbation theory.  We therefore solve the equations of motion \eqref{Maxwell} -- \eqref{Wong} perturbatively in~$g$ but \emph{exactly} in~$\sfa_\mu$. This amounts, of course, to working in background perturbation theory~\cite{DeWitt:1967ub,tHooft:1975uxh,Boulware:1980av,Abbott:1981ke}. Practically, in order to keep track of all factors of $\sfa_\mu$, we expand
\be\label{expand}
	A_\mu \to \frac{1}{g}\,\sfa_\mu + g\, A_\mu + O(g^2) \;, \quad  x \to x + O(g) \;, \quad c^{\sfa} \to c^{\sfa} + O(g) \;,
\ee
plug these into the equations of motion \eqref{Maxwell} -- \eqref{Wong}, and solve order-by-order in $g$. As suggested by \eqref{expand}, to construct the emitted radiation field to leading order in $g$, only the zeroth order orbit and colour charge are required. (At higher orders one can identify e.g.~radiation-reaction on the electron motion~\cite{Krivitsky:1991vt,Higuchi:2002qc,Ilderton:2013dba,DiPiazza:2018luu,Blackburn:2019rfv}.)

\subsection{Zeroth order: charge motion in the background}
To zeroth order in $g$, there is no generated gluon field, so the background remains unchanged and the equations to solve for the particle degrees of freedom are
\bea
\label{Lor1} 
	m\, \ddot{x}_\mu &=& -c^\sfa  f^{\sfa}_{\mu\nu}\, \dot{x}^\nu \,, \\
\label{Wong1} 	\dot{c}^{\sfa}  &=& \im\, {\dot x}_ \mu\, [\sfa^\mu ,\, c]^{\sfa}  \,,
\eea
where $f_{\mu\nu}$ is the field strength \eqref{fmunudef} of $\sfa_\mu$; \eqref{Lor1} and \eqref{Wong1} are just the coupled Lorentz force and Wong's equation in the background.

Now, we use the fact that the background $\sfa_{\mu}$ is given by a Cartan-valued sandwich plane wave \eqref{PGgauge1}. Since such a background is Cartan-valued, only the Cartan-valued components of the colour degrees of freedom enter \eqref{Lor1}. Let $\sfi$ be an index running over the Cartan subalgebra (e.g., for gauge group SU$(N)$, $\sfi=1,\ldots,N-1$), i.e.~$\sfa_\mu \equiv \sfa^\sfi_\mu \sfT^\sfi$ for $\sfT$ the group generators. Then from \eqref{Wong1} it follows that $\dot{c}^{\sfi}=0$. For the remaining components of the colour degrees of freedom, note that $[\sfa^\mu,\, c]^\sfa = (e_k\sfa^\mu) c^\sfa$, where $e_k$ is the root-valued charge of $\sfT^{\sfa}$ with respect to the Cartan-valued background (the meaning of the subscript $k$ will become clear below). Then it follows that $c^{\sfa}(\tau)$ can be expressed as an \emph{abelian} Wilson line:
\bea
\label{c-sol-1}
\dot{c}^\sfa  &=& \im\, e_k\,\sfa^\mu\,{\dot x}_ \mu\,  c^\sfa
\quad
	 \implies 
	 \quad c^\sfa(\tau)  = \exp\left[\im\,e_k\, \int\limits^\tau_{-\infty}\!\ud s\,\sfa^\mu(x(s))\, {\dot x}_ \mu(s)\right]\,  c^\sfa_\text{in} \;,
\eea
where $c_{\text{in}}^{\sfa}:=c^{\sfa}(-\infty)$ are the colour degrees of freedom before the charge enters the sandwich plane wave background. Note that \eqref{c-sol-1} trivially accounts for the constant Cartan components, since the charge $e_k=0$ in the Cartan, and $c^{\sfi}(\tau)=c^{\sfi}_{\text{in}}\equiv e_p^{\sfi}$. Note that it is $e_p^\sfi f^\sfi_{\mu\nu}$ which enters the Lorentz force law \eqref{Lor1}.

Now, the first integral of the Lorentz force \eqref{Lor1} is found by observing that $\ddot{x}^\LCm=0$, so that proper time may be traded for lightfront time $x^\LCm(\tau) = p^\LCm \tau/m$, in which $p^\mu$ denotes the on-shell (i.e.~$p^2=m^2$) free particle momentum before entering the background. In terms of this physical time parameterisation, the solution of the Lorentz force law is given by a time-ordered exponential which truncates at second order due to the nilpotency of $f_{\mu\nu}$; this yields the `background-dressed' kinematic momentum
\be\label{Lor2}
	P_\mu(x^\LCm) = p_\mu - e_p a_\mu(x^\LCm) + \frac{2 p\cdot  e_pa(x^\LCm)-e_p^2\,a^2(x^\LCm)}{2\, p_{+}}\,n_\mu\:,
\ee
where $e_p a_{\perp}:= e_p^{\sfi}\,a^{\sfi}_{\perp}$, with $e_p^{\sfi}$ the (constant) Cartan components of the colour degrees of freedom. Note that, as in \eqref{Lor2}, we often abuse notation and treat the root-valued charges like U$(1)$ charges since the contractions of colour indices are always obvious. The time-dependent momentum $P_\mu(x^\LCm)$ obeys $P^2=m^2$, on-shell\footnote{It is easy to see that $-a_\mu$, which appears in the orbit from direct exponentiation of the field strength, is an alternative gauge potential for $f_{\mu\nu}$. As such $a_\mu$ is directly related to a gauge invariant quantity, and in QED it is convenient to use it as the potential~\cite{Dinu:2012tj} -- this is the analogue of working in Einstein-Rosen coordinates for a gravitational plane wave. Our chosen gauge $\sfa_\mu$ instead corresponds to Brinkmann coordinates which, unlike Einstein-Rosen, have the advantage of being global~\cite{Penrose:1965rx}.} and by direct analogy with classical electrodynamics we see that $e_p$ should be identified as the charge of the particle with respect to the background. Integrating once more, the orbit itself (discarding the irrelevant initial position of the charge) is
\be\label{Lor3}
x^{\mu}(x^-)=\int\limits^{x^-}_{-\infty}\d y\,\frac{P^{\mu}(y)}{p_+}\,.
\ee
This in turn enables the colour degrees of freedom \eqref{c-sol-1} to be expressed in terms of lightfront time rather than proper time:
\be\label{Lor33}
	c^\sfa(x^\LCm)  = \exp\left[\im\,e_k\, x^\mu(x^\LCm)\, a_\mu(x^\LCm) -\im\,e_k\, \int\limits^{x^\LCm}_{-\infty}\!\ud s\, \frac{a(s) \cdot P(s)}{p_\LCp}\right]  c^\sfa_\text{in} \;,
\ee
where the first term in the exponential arises from an integration by parts in \eqref{c-sol-1}.

\subsection{First order: emission and back-reaction}
The charge with orbit $x^\mu$ and colour $c^\sfa$ generates a current which, at order $g$, sources back-reaction on the gluon field $A_\mu$ through the Yang-Mills equations \eqref{Maxwell}. Using the explicit form \eqref{Lor2} -- \eqref{Lor3} of the particle orbit, the current is
\be\label{current}
	J^{\sfa}_\mu(y) = \frac{1}{p_\LCp}\, P_\mu(y^\LCm)\, c^\sfa(y^\LCm) \delta^{\LCperp,\LCp}(y-x(y^\LCm)) \;.
\ee
To order $g$ (but exactly in the background) the Yang-Mills equations become
\be\label{EOM2}
	{D}^2 A^{\sfa}_\mu - {D}_\mu\, {D} \cdot A^{\sfa} +2\im\, e_k\, f_{\mu\nu}\,A^{\sfa\,\nu}= - J^{\sfa}_\mu \;,
\ee
where $A^{\sfa}_{\mu}$ is the emitted gluon field and $D_{\mu}$ is the background covariant derivative.

%
%
Given the form of the background, it is natural to construct the radiation field in lightfront gauge: $n\cdot A = A_\LCp =0$. This also allows for easy comparison with the quantum calculation later.  Since the only nonzero components of the background field strength are $f_{\LCm\LCperp} = -f_{\LCperp\LCm}$, lightfront gauge implies  $\partial_\LCp({D}\cdot A) = J_\LCp$. From this, one obtains a differential equation for the transverse components $A_{\perp}$ and an algebraic equation for $A_{-}$:
\be\label{EOM3}
	{D}^2 A_\LCperp = -J_\LCperp + \partial_\LCperp \frac{J_\LCp}{\partial_\LCp} \;, \qquad \text{and} \qquad 
	A_\LCm = \frac{{\partial}_\LCperp A_\LCperp}{\partial_\LCp} + \frac{J_\LCp}{\partial_\LCp^2} \;,
\ee
where colour indices have been suppressed.

These may be written together covariantly as
\be\label{A-sol}
	A_\mu = -\frac{1}{{D}^2} {\tilde J}_\mu  + \frac{2\,\im}{{D}^2} e_k\,f_{\mu\nu}\, \frac{1}{{D}^2}\, \tilde{J}^\nu  \;,
\ee
in which the gauge-dependent current $\tilde{J}$ is
\be
	{\tilde J}_\mu := J_\mu - {D}_\mu \frac{n\cdot J}{n\cdot \partial}=J_{\mu}-D_{\mu}\,\frac{J_+}{\partial_+} \;.
\ee
Now, we are interested in extracting the emitted \emph{radiation} field at infinity, but \eqref{A-sol} also includes contributions to the (boosted) Coulomb fields of the colour charge. The radiation field can be extracted by projecting, in the limit $x^{-}\to\infty$, onto the basis of free, transverse gluon states of momentum $k_{\mu}$ and polarization $\epsilon_{\mu}$: 
\be\label{BASIS}
	\varphi^\mu(x) := \frac{1}{\sqrt{2k_\LCp\, V}}\, \e^{-\im k\cdot x}\, \epsilon^\mu(k) \;,
\ee
where $k\cdot\epsilon=0$ and the lightfront volume factor $V$ can be set to unity. In lightfront gauge, we have also that $n\cdot\epsilon=0$. Taking the projection we see that 
\be\label{Asympt-Proj}
\begin{split}
	\frac{(\varphi,A)}{(\varphi,\varphi)} &=
	-2k_\LCp\,\int\!\ud^{d-2}x^\LCperp\, \ud x^+ \, \bar{\varphi}^{\mu}(x)\, A_\mu(x) \\ 
	&= 2k_\LCp\,\int\!\ud^{d-2}x^\LCperp\, \ud x^+ \, \bar{\varphi}^\mu(x)\, \int \d^{d}y\, G(x,y)\, \tilde{J}_\mu(y) \;,
\end{split}
\ee
where $G(x,y)$ is the Green's function for $D^2$. So only the first term in \eqref{A-sol} contributes to the radiation field at infinity.

Thus, we only need to solve for $A_{\mu}$ using the first term on the right-hand-side of \eqref{A-sol}. The required inverse of $D^2$ is~\cite{Adamo:2018mpq}:
\be\label{GDEF}
\begin{split}
	G(x,y) &=  \im \e^{\im e_k\,x\cdot a(x^-)} \Bigg[ \int\!\frac{\ud^{d-2} k_\LCperp\,\ud k_\LCp}{(2\pi)^{d-1}\, 2k_\LCp} \, \theta(k_\LCp)\theta(x^\LCm-y^\LCm)\e^{-\im k \cdot (x-y)-\im\int\limits_{y^\LCm}^{x^\LCm} \frac{2e_ka\cdot k- e_k^2a^2}{2k_\LCp}} \Bigg] \e^{-\im e_k\, y\cdot a(y^-)}
\end{split}
\ee
in which the theta-functions arise from imposing retarded boundary conditions. In 
\be\label{clYM1}
A^{\sfa}_{\mu}(x)=-\int \d^{d}y\,G(x,y)\,\tilde{J}^{\sfa}_{\mu}(y)\,,
\ee
an integration by parts moves the derivatives in $\tilde{J}_{\mu}$ onto $G(x,y)$; this means that under the $y^\LCm$ integral in \eqref{clYM1} and the Fourier integral in \eqref{GDEF}, $\tilde{J}_{\mu}$ can be represented by
\be\label{altJtilde}
\tilde{J}^{\sfa}_{\mu}(y) \to \left(\eta_{\mu\nu}-\frac{K_{\mu}(y^-)\,n_{\nu}}{k_+}\right)\,J^{\sfa\,\nu}(y)\,,
\ee
where $K_{\mu}$ is the classical kinematic momentum of the gluon, analogous to $P_{\mu}$ but for momentum $k_{\mu}$ and charge $e_k$, so that $K^2=0$. The $\d^{d-2}y^{\perp}$ and $\d y^{+}$ integrals in \eqref{clYM1} can now be performed trivially against the delta functions appearing in the current \eqref{current}; a short calculation leaves:
\begin{multline}\label{clYM2}
	-\im\,c^{\sfa}_{\text{in}}\,\e^{\im e_k\,x\cdot a(x^-)}\int \!\frac{\d^{d-2}k_\LCperp\,\d k_\LCp}{(2\pi)^{d-1}\, 2k_\LCp}\,\theta(k_+)\,\int\limits_{-\infty}^{x^-}\d y^{-}\left(\eta_{\mu\nu}-\frac{K_{\mu}(y^-)\,n_{\nu}}{k_+}\right)\,\frac{P^{\nu}(y^-)}{p_+} \\
\times \exp\left[-\im\,k\cdot x-\im\int\limits_{-\infty}^{x^-}\d s\,\frac{2e_k\,k\cdot a(s)-e_k^2\,a^2(s)}{2\,k_+}+\im\int_{-\infty}^{y^-}\!\d s\,\frac{P\cdot K(s)}{p_+}\right]\,,
\end{multline}
using the solutions \eqref{Lor3} and \eqref{Lor33}. Now, to extract the classical radiation field at infinity we use boundary conditions\footnote{In general, $a_{\perp}(+\infty)$ can be a non-vanishing constant; this corresponds to the gauge-theoretic memory effect~\cite{Dinu:2012tj,Bieri:2013hqa,Pate:2017vwa}. Accounting for this more general boundary condition is straightforward, but requires additional gauge terms in \eqref{BASIS} to account for the fact that the potential does not vanish asymptotically~\cite{Kibble:1965zza}.} $a_{\perp}(+\infty)=0$ to find: 
\be\begin{split}\label{Classical-result*}
\mathcal{A}^{\sfa}(k) :=&\lim_{x^\LCm\to\infty } -2k_\LCp\,\int\!\ud^{d-2}x^\LCperp\, \ud x^+ \, \bar{\varphi}^\mu(x)  \frac{1}{{D}^2} \tilde{J}_\mu(x)   \\
&= \frac{\im c^\sfa_\text{in}}{\sqrt{2\,k_\LCp}}
\,
	\epsilon^\mu\,  \int\limits_{-\infty}^{+\infty}\ud y^\LCm \left(\eta_{\mu\nu} - \frac{K_\mu(y^\LCm)\, n_\nu}{k_\LCp}\right) \frac{P^\nu(y^\LCm)}{p_\LCp}\, \exp\left[ \im\, \int\limits^{y^\LCm}_{-\infty} \!\ud s\, \frac{P\cdot K(s)}{p_\LCp} \right] \;,
\end{split}
\ee
dropping an irrelevant overall phase.

We note that the tensor part of \eqref{Classical-result*} can be rewritten as
\be\label{Proj}
	\left(\eta_{\mu\nu} - \frac{K_\mu(y^\LCm)\, n_\nu + n_\mu\, K_\nu(y^\LCm)}{k_\LCp}\right)\,P^\nu(y^\LCm):=\mathbb{P}_{\mu\nu}\,P^{\nu}(y^-)\,,
\ee
since the additional (third) term in $\mathbb{P}_{\mu\nu}$ generates only a total derivative in \eqref{Classical-result*}. The projector $\mathbb{P}_{\mu\nu}$ is `doubly' transverse, to both $K_\mu$ and $n_\mu$. Thus, the classical back-reaction gluon field at infinity can be written as: 
\be\label{Classical-result}
\mathcal{A}^{\sfa}(k)=\frac{\im c^\sfa_\text{in}}{\sqrt{2\,k_\LCp}}\,\epsilon^{\mu}\,\int\limits_{-\infty}^{+\infty}\ud y^\LCm \,\mathbb{P}_{\mu\nu}\, \frac{P^\nu}{p_\LCp}(y^-)\, \exp\left[ \im\, \int\limits^{y^\LCm}_{-\infty} \!\ud s\, \frac{P\cdot K(s)}{p_\LCp} \right] \,.
\ee

\subsubsection*{Fundamental matter}

Note that if the colour charge was valued in the fundamental (rather than adjoint) of the gauge group, the relevant classical equation of motion for the colour degrees of freedom is
\be
	\dot \theta = \im\, g\, {\dot x}^\mu\, A_\mu \cdot \theta \,,
\ee
where $\theta(\tau)$ is a fundamental vector. In this case, $c^\sfa(\tau) := \theta^\dagger \cdot \sfT^\sfa \cdot \theta$ enters just as in the adjoint-valued case, and \eqref{Wong} still governs its evolution~\cite{Balachandran1977}. Hence the results for the adjoint case transfer trivially to the case of fundamental matter.

\section{Classical gravitational mass dynamics}
\label{SECT:GR}

In this section we construct, in analogy to the Yang-Mills calculation above, the leading order emitted graviton radiation from a massive scalar particle crossing a gravitational plane wave spacetime. A single massive particle coupled to classical gravity is described by the worldline theory for a point particle\footnote{Allowing for finite-size effects, the point particle theory is still an effective description in the low velocity approximation up to sixth order in $v/c$~\cite{Goldberger:2004jt}.} coupled to the Einstein-Hilbert action. The position of the particle as a function of proper time $x^{\mu}(\tau)$ is governed by the geodesic equation 
\be\label{geodesic}
\ddot{x}^{\mu}=-\Gamma_{\nu\rho}^{\mu}\,\dot{x}^{\nu}\,\dot{x}^{\rho}\,,
\ee
where $\Gamma_{\nu\rho}^{\mu}$ are the Christoffel symbols of the Levi-Civita connection determined by the Einstein equations:
\be\label{Einstein}
R_{\mu\nu}-\frac{1}{2}\,R\,g_{\mu\nu}=\kappa\,T_{\mu\nu}\,,
\ee
where $\kappa$ is the gravitational coupling constant, $R_{\mu\nu}$ is the Ricci tensor, $R$ the Ricci scalar and
\be\label{SEtens}
T_{\mu\nu}(x)=m\,\int\ud\tau\,\dot{x}^{\mu}(\tau)\,\dot{x}^{\nu}(\tau)\,\delta^{d}(x-x(\tau))\,,
\ee
is the stress-energy tensor of the scalar point particle.

This massive particle interacts with a given background metric $g_{\text{in}\mu\nu}\equiv g_{\mu\nu}$, which is initially present and obeys the vacuum Einstein equations. This background accelerates the particle which emits gravitational radiation, producing back-reaction on $g_{\mu\nu}$. We allow the gravitational background to be arbitrarily strong, so $\kappa\,g_{\mu\nu}$ is characterised by a dimensionless coupling greater than one and cannot be treated perturbatively. Thus, we solve \eqref{geodesic}, \eqref{Einstein} perturbatively in $\kappa$ but \emph{exactly} in $g_{\mu\nu}$, as in background perturbation theory. Following the gauge theory computations in Section~\ref{SECT:YM}, we expand
\be\label{grexpand}
g_{\mu\nu}\to\frac{1}{\kappa}\,g_{\mu\nu}+\kappa\,h_{\mu\nu}+O(\kappa^2)\,, \qquad x\to x +O(\kappa)\,,
\ee
with only the displayed terms required to construct the emitted radiation field to leading order in $\kappa$. 


\subsection{Zeroth order: particle motion in the background}

At zeroth order in $\kappa$, there is no generated graviton field and one is left to consider the motion of the massive particle in the background metric, as governed by the geodesic equation \eqref{geodesic}. On a general background, this equation admits a formal solution for the particle's kinematic momentum $P^{\mu}(\tau)$ in terms of the particle momentum prior to entering the background, $p^{\mu}$, and a gravitational Wilson line (e.g., \cite{Modanese:1993zh,Brandhuber:2008tf,Goldberger:2016iau}):
\be\label{GRWilson}
P^{\mu}(\tau)=W^{\mu}_{\nu}(\tau)\,p^{\nu}\,, \qquad W^{\mu}_{\nu}(\tau):=\mathcal{P}\exp\left[-\int_{-\infty}^{\tau}\d s\, \Gamma^{\mu}_{\nu\rho}\,\dot{x}^{\rho}(s)\right]\,,
\ee
which depends implicitly on $x^{\mu}(\tau)$, with $\mathcal{P}$ denoting path-ordering. However, when $g_{\mu\nu}$ is a sandwich plane wave \eqref{grPW1}, the simplicity of the metric ensures that the path-ordered expansion of the gravitational Wilson line truncates at low order.

In particular, the only non-vanishing Christoffel symbols of the plane wave background are:
\be\label{Christoffel}
\Gamma_{--}^{a}=-H^{a}_{b}(x^-)\,x^{b}\,, \qquad \Gamma_{--}^{+}=-\frac{1}{2}\dot{H}_{ab}(x^-)\,x^{a}x^{b}\,, \qquad \Gamma_{-a}^{+}=-H_{ab}(x^-)\,x^{b}\,.
\ee
From this, it immediately follows that $\ddot{x}^{-}=0$, so that $x^-$ can be substituted for proper time through $x^{-}=p^{-}\tau/m$, and the remaining components of the kinematic momentum are
\be\label{grMom}
P_{a}(x^-)=p_{i}\,E^{i}_{a}(x^-)+p_{+}\sigma_{ab}(x^-)\,x^{b}(x^-)\,,
\ee
\begin{equation*}
P_{-}(x^-)=\frac{m^2}{2\,p_+}+\gamma^{ij}(x^-)\,\frac{p_{i}\,p_{j}}{2\,p_+}+\frac{p_+}{2}\dot{\sigma}_{bc}(x^-)\,x^{b}(x^-)\,x^{c}(x^-)+p_{i}\,\dot{E}^{i}_{b}(x^-)\,x^{b}(x^-)\,,
\end{equation*}
where $E^{i}_{a}$, $\gamma_{ij}$ and $\sigma_{ab}$ are defined by \eqref{vielbein}, \eqref{transmet} and \eqref{deftensor}, respectively. It is easy to see that $P^{2}=g^{\mu\nu}P_{\mu}P_{\nu}=m^2$ on-shell. Note that the kinematic momentum depends implicitly on the solution $x^{a}(x^-)$ for the particle trajectory in the transverse plane, determined by the time-dependent harmonic oscillator equations
\be\label{oscillator}
\ddot{x}^{a}=\frac{p_+^2}{m^2}\,H^{a}_{b}(x^-)\,x^{b}(x^-)\,.
\ee
Of course, the trajectory can still be formally expressed in terms of the kinematic momentum as
\be\label{trajectory}
x^{\mu}(x^-)=\int\limits_{-\infty}^{x^-}\d y\,\frac{P^{\mu}(y)}{p_+}\,.
\ee


\subsection{First order: emission and back-reaction}

The particle's motion in the plane wave background generates a stress-energy tensor
\be\label{grsource}
T^{\mu\nu}(y)=\frac{P^{\mu}(y^-)\,P^{\nu}(y^-)}{p_+}\,\delta^{+,\perp}(y-x(y^-))\,,
\ee
which in turn will source graviton emission through the Einstein equation \eqref{Einstein}. Following the symmetries of the plane wave background, it is convenient to impose lightfront and traceless gauge conditions on the graviton field: $n^{\mu}h_{\mu\nu}=0=h^{\mu}_{\mu}$. With this choice, the Einstein equation at first order in $\kappa$ (but exactly in $g_{\mu\nu}$) becomes:
\be\label{Einstein1} 
\nabla^{2}h_{\mu\nu}-2\,\nabla_{(\mu}\nabla^{\rho}h_{\nu)\rho}+g_{\mu\nu}\,\nabla^{\rho}\nabla^{\sigma}h_{\rho\sigma}-2\,n_{\mu}n_{\nu}\,h^{ab}\,H_{ab}=-2\,T_{\mu\nu}\,,
\ee
where $\nabla_{\mu}$ is the covariant derivative of the background and all indices are raised an lowered with $g_{\mu\nu}$. 

Using the lightfront condition ($h_{+\mu}=0$), it follows that
\be\label{lightfront1}
\nabla^{\rho}h_{\mu\rho}=2\,\frac{T_{+\mu}}{\partial_+}+n_{\mu}\,\frac{\nabla^{\rho}\nabla^{\sigma}h_{\rho\sigma}}{\partial_+}\,,
\ee
which enables \eqref{Einstein1} to be rewritten as
\be\label{Einstein2}
h_{\mu\nu}=-\frac{2}{\nabla^2}\,\widetilde{T}_{\mu\nu}+\frac{2}{\nabla^2}\,n_{\mu}n_{\nu}\,h^{ab}\,H_{ab}\,,
\ee
where the gauge-dependent stress-energy tensor is:
\be\label{modstress}
\widetilde{T}_{\mu\nu}:=T_{\mu\nu}-2\,\frac{\nabla_{(\mu}T_{\nu)+}}{\partial_+}+\nabla_{(\mu}\nabla_{\nu)}\frac{T_{++}}{\partial_+^2}\,.
\ee
The second term on the right-hand-side of \eqref{Einstein2} arises from the Riemann curvature tensor of the plane wave background, and can be expressed in terms of a further inverse of $\nabla^2$ acting on components of $T_{\mu\nu}$.

Recall that we are interested in the emitted graviton \emph{radiation} field at infinity, and \eqref{Einstein2} also encodes non-radiative contributions (e.g., the Schwarzschild fields associated to the point mass). The radiative contribution is singled out by projecting onto the basis of free, transverse graviton states of momentum $k_\mu$ and polarization $\epsilon_{\mu\nu}$:
\be\label{grBASIS}
	\varphi^{\mu\nu}(x) := \frac{1}{\sqrt{2k_\LCp\, V}}\, \e^{-\im k\cdot x}\, \epsilon^{\mu\nu}(k) \;,
\ee
in the limit $x^-\to\infty$. Here, the lightfront volume $V$ will be set to unity, and transverse lightfront gauge sets $k^{\mu}\epsilon_{\mu\nu}=0=n^{\mu}\epsilon_{\mu\nu}$. Projection onto these asymptotic states gives 
\be\begin{split}\label{Asympt-Proj}
	\frac{(\varphi,h)}{(\varphi,\varphi)} &=2k_\LCp\,\int\!\ud^{d-2}x^\LCperp\, \ud x^\LCp \, \bar{\varphi}^{\mu\nu}(x)\, h_{\mu\nu}(x) \\
	&= -4k_\LCp\,\int\!\ud^{d-2}x^\LCperp\, \ud x^\LCp \, \bar{\varphi}^{\mu\nu}(x) \int \d^{d}y\, G(x,y)\, \tilde{T}_{\mu\nu}(y) \;,
\end{split}
\ee
where $G(x,y)$ is the Green's function for the Laplacian $\nabla^2$. Only the first term on the right-hand-side of \eqref{Einstein2} contributes to the graviton radiation field at infinity.

So we can consider the Einstein equation \eqref{Einstein2} with only the first term on the right-hand-side. The inverse of $\nabla^2$ is~\cite{Gibbons:1975jb,Ward:1987ws}:
\begin{multline}\label{LapGF}
G(x,y)=\frac{1}{\sqrt{|E(x)|}}\int \frac{\d^{d-2} k_{\perp}\,\d^{2}k_{+}}{(2\pi)^{d-1}\, 2k_\LCp}\, \theta(k_\LCp)\theta(x^\LCm-y^\LCm)\frac{1}{\sqrt{|E(y)|}} \\
\exp\left[ -\im\,\frac{k_{i}k_{j}}{2\,k_+}\int_{y^-}^{x^{-}}\d s\,\gamma^{ij}(s) -\im\, k_{+}(x-y)^+ -\im\,k_{i}(E^{i}_{a}(x) x^{a}-E^{i}_{a}(y) y^a)\right. \\
-\im\frac{k_+}{2}(\sigma_{ab}(x)x^{a}x^{b}-\sigma_{ab}(y)y^{a}y^{b})\Bigg]\,,
\end{multline}
and in the expression
\be\label{clGR1}
h_{\mu\nu}(x)=-2\,\int \d^{d}y\,G(x,y)\,\widetilde{T}_{\mu\nu}(y)\,,
\ee
a series of integrations by parts enables us to move the derivatives in $\tilde{T}_{\mu\nu}$ onto $G(x,y)$. So inside the Fourier integral in \eqref{LapGF}--\eqref{clGR1}, $\widetilde{T}_{\mu\nu}$ can be replaced by:
\begin{multline}\label{altTtilde}
\widetilde{T}_{\mu\nu}(y)\to\left(g_{\sigma(\mu}\,g_{\nu)\rho}-\frac{2}{k_\LCp} g_{\sigma(\mu} K_{\nu)}(y)\,n_{\rho}+n_{\sigma}n_{\rho}\,\frac{K_{(\mu}(y)\,K_{\nu)}(y)}{k_+^2}\right. \\
\left. -\frac{\im}{k_+}n_{\sigma}n_{\rho}\,\sigma_{ab}(y)\,\delta^{a}_{(\mu}\delta^{b}_{\nu)}\right) T^{\sigma\rho}(y)\,,
\end{multline}
where the kinematic graviton momentum $K_{\mu}$ obeys $K^2=0$. The final term appearing in the parentheses, proportional to $\sigma_{ab}$, arises due to a cross term in the integration by parts of the form $\nabla_{(\mu}K_{\nu)}=k_+\,\sigma_{ab}\delta^{a}_{(\mu}\delta^{b}_{\nu)}$. This tensor structure differs only by terms which are total derivatives from
\be\label{altTtilde*}
	\widetilde{T}_{\mu\nu}(y) = \left(\mathbb{P}_{\mu\lambda}\,\mathbb{P}_{\nu\sigma}-\frac{\im}{k_+}\,n_{\mu}\,n_{\nu}\,\delta^{a}_{\lambda}\delta^{b}_{\sigma}\,\sigma_{ab}\right) T^{\sigma\rho}(y)
	=:\mathbb{P}_{\mu\nu\sigma\rho}\,T^{\sigma\rho}(y)\,,
\ee
in which $\mathbb{P}_{\mu\nu}$ is itself the natural analogue of the projector \eqref{Proj}: 
\be\label{grpol1*}
\mathbb{P}_{\mu\nu} = g_{\mu\nu}-\frac{K_{\mu}\,n_{\nu}+n_{\mu}\,K_{\nu}}{k_{+}} \;.
\ee
We thus use \eqref{altTtilde*} from here on. 

Returning to \eqref{clGR1}, the integrals in $\d^{d-2}y^{\perp}$ and $\d y^{+}$ can be performed against the delta functions appearing in the stress-energy tensor \eqref{grsource} to give:
\be\label{clGR2}
-\frac{2}{\sqrt{|E(x)|}}\,\int\!\frac{\d^{d-2}k_\LCperp\,\d k_\LCp}{(2\pi)^{d-1}\, 2k_\LCp}\,\theta(k_+)\,\int\limits_{-\infty}^{x^-}\frac{\d y^{-}}{\sqrt{|E(y)|}}\,\mathbb{P}_{\mu\nu\sigma\rho}\,\frac{P^{\sigma}\,P^{\rho}}{p_+}(y^-) \,\e^{-\im\mathcal{F}(x,y^-)}\,.
\ee
The argument of the exponential can be written
\be\label{GRexp1}
\mathcal{F}(x,y^-):=k_{+}\,\left(X-X(y^-)\right)^{+}+k_{i}\left(X-X(y^-)\right)^{i}+\frac{k_{i}\,k_{j}}{2\,k_{+}}\int_{y^-}^{x^{-}}\ud s\,\gamma^{ij}(s)\,,
\ee
where the coordinates $X^{+}$, $X^{i}$ are defined through the diffeomorphism
\be\label{ERdiffeo}
X^{+}:=x^{+}+\frac{1}{2}\dot{E}^{i}_{a}\,E_{b\,i}\,x^{a}x^{b}\,, \qquad X^{i}:=E^{i}_{a}\,x^{a}\,.
\ee
Using \eqref{trajectory} and the boundary conditions\footnote{While $\sigma_{ab}$ generically falls off as $1/x^{-}$ as $x^{-}\to\infty$, the vielbein $E^{i}_{a}$ can, in general, approach an arbitrary constant matrix. This encodes the memory effect of the plane wave background~\cite{Bieri:2013ada,Zhang:2017rno,Adamo:2017nia}, and can be accounted for by modifying the basis \eqref{grBASIS} with additional gauge terms.}
\be\label{Asympt}
\lim_{x^{-}\rightarrow\infty}E^{i}_{a}(x^-)=\delta^{i}_{a}\,, \qquad \lim_{x^{-}\rightarrow\infty}\sigma_{ab}(x^-)=0\,,
\ee
the classical radiation field at infinity is:
\be\begin{split}\label{clGR3}
\mathcal{H}(k)&:=\lim_{x^{-}\to\infty}-4k_{+}\,\int\!\d^{d-2}x^\perp\,\d x^{+}\,\bar{\varphi}^{\mu\nu}(x)\,\frac{1}{\nabla^2}\widetilde{T}_{\mu\nu}(x) \\
&= \frac{-2}{\sqrt{2\,k_+}}\,\epsilon^{\mu\nu}\,\int\limits_{-\infty}^{+\infty}\d y^{-}\,\mathbb{P}_{\mu\nu\sigma\rho}\,\frac{P^{\sigma}\,P^{\rho}}{p_+}(y^-)\,\exp\left[\im\,\int\limits^{y^\LCm}_{-\infty} \!\ud s\,\frac{P\cdot K(s)}{p_\LCp} \right]\,,
\end{split}
\ee
where an irrelevant overall phase has been dropped.

\section{Non-linear Compton scattering}\label{NLComp}
\begin{figure}[t!!]
\centering\includegraphics[width=0.2\textwidth]{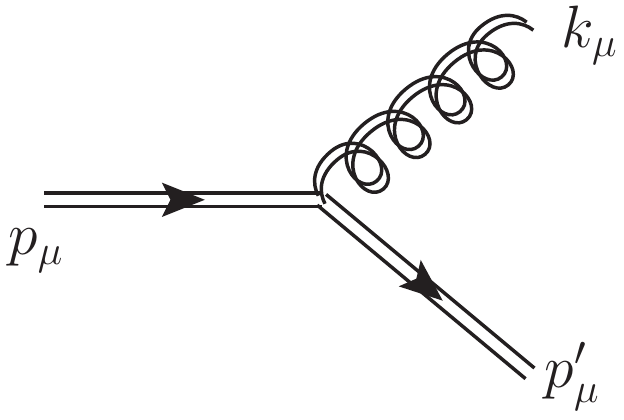}
	\caption{\label{FIG:TREE} Tree-level non-linear Compton scattering on a plane wave background. Double lines indicate that the particles are fully dressed by the plane wave: in vacuum, the corresponding amplitudes would vanish by momentum conservation, but here they are nontrivial functions of the scattering momenta and the background.}
\end{figure}

In perturbative QFT, back-reaction on a (here plane wave) background corresponds -- at lowest order -- to the 3-point tree-level scattering amplitude of two charged/massive particles (one incoming, one outgoing) and one emitted gauge boson, see Fig.~\ref{FIG:TREE}. Such 3-point amplitudes are referred to as \emph{non-linear Compton} scattering amplitudes~\cite{nikishov64} for reasons we will make clear below. Here we compute and analyse these amplitudes for gluon and graviton emission in gauge theory and gravity, respectively, with massive scalar legs.

\subsection{Gluon emission amplitude}
Perturbative quantum field theory of colour-charged fields around a gauge theory plane wave background can be studied explicitly thanks to the high degree of symmetry in play. `Free' fields are characterised by the same quantum numbers as on a trivial background. For $k_{\mu}$ an on-shell (massless or massive) momentum, define the scalar function~\cite{volkov35,Seipt:2017ckc,Adamo:2017nia}
\be\label{gaugeHamJac}
\phi_{k}=k\cdot x -e\,a_{\perp}(x^{-})\,x^{\perp}+\int^{x^{-}}\d s\,\frac{2e\,a(s)\cdot k-e^{2}\,a^{2}(s)}{2\,k_{+}}\,,
\ee
in which $e$ is the charge of the free field with respect to the Cartan-valued background. This $\phi_k$ is nothing but the solution to the classical Hamilton-Jacobi equations. A massive adjoint valued scalar $\Phi^{\sfa}$ obeying $\left(D^2+m^2\right) \Phi=0$, for $D_{\mu}=\partial_{\mu}-\im\,e\,\mathsf{a}_{\mu}$ the background covariant derivative, may be represented as
\be\label{mascal}
\Phi^{\sfa}(x)=\sfT^{\sfa}\,\e^{\pm\im\,\phi_{k}}\,, 
\ee
where $\sfT^{\sfa}$ is a generator of the gauge group, $k^2=m^2$ is the momentum before entering the plane wave background (i.e., as $x^{-}\rightarrow-\infty$) and the sign in the exponential dictates whether the scalar is incoming ($-$) or outgoing ($+$). This is analogous to the usual momentum eigenstate representation; defining $K_{\mu}:=e^{i\phi_k}\, \im D_{\mu} e^{-i\phi_{k}}$, we find
\be
	K_{\mu}  = k_\mu - e a_\mu + n_\mu 
	\frac{2e\,a(s)\cdot k-e^{2}\,a^{2}(s)}{2\,k_{+}}\,,
\ee
which recovers the kinematic momentum of a particle (with initial momentum $k_\mu$ and charge $e$) traversing the plane wave background, as we found in the classical theory. It is on-shell, $K^2=k^2$, by virtue of the fact that $\phi_k$ solves the Hamilton-Jacobi equations.

Similarly, a gluon perturbation $A^{\sfa}_{\mu}$ in the plane wave background is given by
\be\label{gluonpert}
A^{\sfa}_{\mu}(x)=\sfT^{\sfa}\,\cE_{\mu}(x^-)\,\e^{\pm\im\,\phi_k}\,,
\ee
where $k^2=0$ and $\cE_{\mu}(x^-)$ is a dressed polarization vector which in lightfront-Feynman gauge ($n\cdot A=0=D\cdot A$) can be expressed in terms of the free, un-dressed polarization vector $\epsilon_{\nu}(k)$ using the doubly-transverse projector $\mathbb{P}_{\mu\nu}$ from \eqref{Proj}:
\be\label{glpol}
\cE_{\mu}= 
\mathbb{P}_{\mu\nu}\,\epsilon^{\nu}\,.
\ee
It is easy to see that $n\cdot\cE=0=K\cdot\cE$, so that this dressed polarization is transverse to the dressed momentum. It is straightforward to generalize these constructions to free fields of other spins, or valued in other representations of the gauge group.


We now turn to the tree-level amplitude for an incoming colour-charged massive scalar to emit a gluon upon crossing a gauge theory plane wave background. This amplitude is computed by the cubic part of the background field Lagrangian for a charged scalar coupled to Yang-Mills theory in the plane wave background:
\be\label{YMs1}
 g\,\int \ud^{d}x\; \tr\left([A_{\mu},\,\Phi]\,D^{\mu}\Phi\right)\,,
\ee
evaluated using the wavefunctions \eqref{mascal} and \eqref{gluonpert}. Let the incoming and outgoing scalars have momenta $p_{\mu}$ (obeying $p^2=m^2$) and $p'_{\mu}$ (obeying $p^{\prime\,2}=m^2$) respectively, and the emitted gluon have momentum $k_{\mu}$ (obeying $k^2=0$) and polarization $\epsilon_\mu$ (obeying $k\cdot\epsilon=0$).

Charge conservation implies that the root-valued adjoint charges for the three particles obey $e_{p}=e_{p'}+e_k$, and on the support of this relation the position space integrals in $x^+$ and $x^\perp$ can be performed straightforwardly to give:
\be\label{YMs2}
\im\,g\,\mathrm{f}^{\mathsf{abc}}\,(2\pi)^{d-1}\,\delta^{+,\perp}(p'+k-p)\,\int\limits_{-\infty}^{+\infty}\d x^{-}\,\cE\cdot(P+P')(x^-)\, \exp\left[\im \int\limits^{x^{\LCm}}\!\d s\, \frac{P\cdot K(s)}{(p-k)_{+}}\right]\,,
\ee
where the indices on $\mathrm{f}^{\mathsf{abc}}$ corresponding to each of the three adjoint-valued particles, and the $d-1$ delta functions enforce momentum conservation (of the un-dressed momenta) in the $x^+$ and $x^\perp$ directions. This reduced momentum conservation implies that
\be\label{4Vrel}
	P'_\mu(x^\LCm) = P_\mu(x^\LCm) - K_\mu(x^\LCm) + \frac{K\cdot P(x^\LCm)}{(p-k)_{+}}\,n_\mu \,,
\ee
which leads to the simple form of the exponential factor. From \eqref{4Vrel} it also follows that
\be\label{polarcons}
 \cE\cdot(P+P')(x^-)=2\cE\cdot P(x^-)-\cE\cdot K(x^-)=2\cE\cdot P(x^-)\,,
\ee
with the final equality following since the dressed gluon polarization is on-shell with respect to the dressed gluon momentum, $\cE\cdot K=0$, see \eqref{glpol}. Thus the tree-level non-linear Compton amplitude for gluon emission from a colour-charged scalar is given by 
\be\label{YMs3}
2\im\,g\,\mathrm{f}^{\mathsf{abc}}\,(2\pi)^{d-1}\,\delta^{+,\perp}(p'+k-p)\,\int\limits_{-\infty}^{+\infty}\d x^{-}\,\cE\cdot P(x^-)\,\e^{\im\, \cV[p,k]}\,,
\ee
where we define the `\emph{Volkov exponent}' for gluon emission by
\be\label{glVolkov}
\cV[p,k]:=\int\limits^{x^{\LCm}}\!\d s\, \frac{P\cdot K(s)}{(p-k)_{+}}\,.
\ee
Such quantities are universal for three-point amplitudes on plane wave backgrounds (cf., \cite{Ilderton:2013dba,Seipt:2017ckc,Adamo:2019zmk}).

\subsubsection*{Classical limit}

The classical limit of \eqref{YMs3} should encode the back-reaction computed from classical colour-charge dynamics in Sect.~\ref{SECT:YM}. In general, taking the classical limit of an observable in QFT requires careful re-introduction of factors of $\hbar$ followed by the $\hbar\to0$ limit~\cite{Kosower:2018adc}. However, for non-linear Compton scattering (which is given by a tree-level contact diagram in background perturbation theory), the classical limit can be implemented straightforwardly without the need to explicitly re-introduce powers of $\hbar$.  

To make the comparison between the two calculations explicit, project the free colour-indices of the scattered scalar onto some adjoint basis, with the identification:
\be\label{sclim1}
c^{\sfa}_{\mathrm{in}}\equiv\mathrm{f}^{\mathsf{abc}}\,\mathsf{X}^{\mathsf{b}}\,\mathsf{Y}^{\mathsf{c}}\,,
\ee
and define the non-linear Compton amplitude stripped of momentum-conserving delta functions
\be\label{sclim2}
\mathcal{M}_{3}^{\mathsf{a}}:=\im g\,\frac{2\,c^{\sfa}_{\mathrm{in}}}{\sqrt{2k_\LCp \, 2p_\LCp\, 2(p-k)_\LCp}}\,\int\limits_{-\infty}^{+\infty}\d x^{-}\,\cE\cdot P(x^-)\,\e^{\im\, \cV[p,k]}\,,
\ee
in which the prefactor includes the correct single-particle state normalisations. Taking the classical limit amounts to assuming that the emitted gluon momentum is negligible compared to that of the incoming scalar, $k_\LCp \ll p_\LCp$.  In the Volkov exponent $\mathcal{V}$ this assumption means replacing $(p-k)^{-1}_{+} \to p^{-1}_{+}$, which immediately recovers the exponent in the classical radiation field \eqref{Classical-result}. In the same limit, the numerical and normalisation factors in (\ref{sclim2}) combine to $1/(p_\LCp\sqrt{2k_\LCp})$, and we obtain
\be\label{sclim3}
\mathcal{M}_{3}^{\mathsf{a}}\xrightarrow{\mathrm{classical}} \mathcal{A}^\sfa (k) \;.
\ee
That is, we recover the radiation field at infinity \eqref{Classical-result} generated by classical back-reaction. This is natural: the quantum amplitude is the projection of the time-evolved state onto a (one scalar plus) single-gluon state, while $\mathcal{A}^{\sfa}$ is the projection of the time-evolved gluon field onto the corresponding basis of single-gluon (also plane wave) radiation fields.



There is one subtlety in comparing the classical and quantum calculations: in the scattering amplitude, the momentum conserving delta-function limits the gluon momentum to $0<k_\LCp < p_\LCp$, whereas in the classical result \eqref{Classical-result} there is no such restriction. The assumption that $k_\LCp \ll p_\LCp$ means effectively ignoring, in the computation of classical observables, the upper limit appearing on any $k_\LCp$ integrals. We remark that the observation of quantum deviations from classical predictions in the equivalent QED setup of radiation emitted from electrons scattered off a background (plane-wave-like) laser has only recently been observed experimentally~\cite{Cole:2017zca,Poder:2018ifi}  (see also~\cite{Wistisen:2017pgr}.)

\subsubsection*{Perturbative limit}
It is instructive to also consider the perturbative limit of \eqref{YMs3} to see why the nomenclature of `non-linear Compton scattering' is appropriate.

The perturbative limit is given by expanding the amplitude from \eqref{YMs1} in powers of $a_\perp$, keeping only those terms which are linear. The zeroth order terms, which are independent of the background, give zero by momentum conservation. At linear order, some algebra shows that one recovers the standard amplitude for Compton scattering, on a trivial background, of two scalars (momenta $p$, $p'$ and mass $m$) and two gluons: one outgoing with momentum and polarization $\{k,\epsilon\}$ and the other incoming with momentum and polarisation $\{q,\widetilde{a}\}$, in which $\widetilde{a}_{\mu}(\omega)$ is the Fourier transform of the background, and serves as the polarization degrees of freedom ($q\cdot \widetilde{a} =0$).

To see how this comes about, it is revealing to go back a step and expand each of the contributing dressed wavefunctions in turn, rather than the amplitude as a whole. Consider first the incoming scalar wavefunction; expanding up to linear order
\be\label{pertlim1}
\e^{-\im\,\phi_{p}}=\e^{-\im\,p\cdot x}\left[1 -\im\,\frac{e_{p}}{p_+}\int^{x^-}\!\!\d s\, p\cdot a(s)+\cdots\right]\,.
\ee
We rewrite this by Fourier transforming the background degrees of freedom as 
\be\label{bFourier}
a_{\perp}(x^-)=\int\frac{\d\omega}{2\pi}\,\e^{-\im\,\omega x^{-}}\,\widetilde{a}_{\perp}(\omega)\,,
\ee
and defining the massless momentum $q_{\mu}:=\omega\,n_{\mu}$ using the Fourier frequency; this gives
\be\label{pertlim2}
\e^{-\im\,\phi_{p}}=\e^{-\im\,p\cdot x}\left[1+ e_{p}\,\int\frac{\d\omega}{2\pi}\,\e^{-\im\,q\cdot x}\,\frac{\widetilde{a}\cdot p}{q\cdot p}+\cdots\right]\,,
\ee
where terms which do not contribute due to charge conservation have been dropped. Acting with the background covariant derivative appearing in \eqref{YMs1}, the linear terms are
\be\label{pertlim2D}
	\im D_\mu e^{-\im \phi_p} = p_\mu e^{\im p.x} + e_p \int\!\frac{\ud \omega}{2\pi}\, e^{-i(p+q)\cdot x} \bigg[ p_\mu \frac{\widetilde a\cdot p}{q\cdot p} -\widetilde a_\mu +  q_\mu\frac{\widetilde a\cdot p}{q\cdot p} \bigg] \;.
\ee
In the perturbative limit, the charge $e_p$ is proportional to another power of the coupling $g$, hence the amplitude is proportional to $g^2$ multiplied by an appropriate colour factor. Inserting the linear term of \eqref{pertlim2D} into \eqref{YMs1} and ignoring overall factors, along with the $\d\omega$ integral, one has:
\be\label{pertlim3}
	\epsilon\cdot p\frac{\widetilde{a}\cdot p\,}{q\cdot p} - \epsilon\cdot \widetilde a=2\,\epsilon\cdot p \frac{\widetilde{a}\cdot p}{(q+p)^2-m^2}\, - \epsilon\cdot \widetilde a\,,
\ee
using momentum conservation and the fact that the polarization $\epsilon_{\mu}$ obeys $q\cdot\epsilon=0=k\cdot\epsilon$. The first term is precisely the $s$-channel contribution to Compton scattering. The second term in \eqref{pertlim3} gives a contribution to the gluon-scalar seagull vertex. A similar expansion of the outgoing scalar wavefunction is easily seen to encode the $u$-channel contribution to Compton scattering along with the remaining contribution to the seagull.

Finally, the expansion of the gluon wavefunction receives contributions from both the exponential and the dressed polarization:
\be\label{pertlim4}
\e^{\im\,k\cdot x}\,\epsilon_\mu +e_{k}\,\int\frac{\d \omega}{2\pi}\, \e^{\im(k-q)\cdot x} \left(\frac{q_\mu \, \widetilde{a}\cdot\epsilon - \epsilon_\mu \widetilde{a}\cdot k }{q\cdot k}\right)\,.
\ee 
Inserting this into \eqref{YMs1} gives
\be
	\bigg(\frac{q_\mu \, \widetilde{a}\cdot\epsilon - \epsilon_\mu\, \widetilde{a}\cdot k }{q\cdot k}\bigg)\,(p+p')^\mu = 2\,\frac{q\cdot (p+p')\,\widetilde a \cdot \epsilon - \epsilon\cdot (p+p')\, \widetilde a\cdot k}{(q+k)^2}\,,
\ee
which is the cubic gluon vertex contribution to Compton scattering (in lightfront gauge).

Thus, the perturbative limit of non-linear Compton scattering does indeed reproduce the amplitude for ordinary Compton scattering on a trivial background, convoluted with the frequency distribution of photons coming from the plane wave. This holds for all non-linear Compton amplitudes considered in this section, following similar calculations.

\subsection{Graviton emission amplitude}

We now turn to the tree-level amplitude corresponding to graviton emission from a massive scalar crossing a plane wave space-time, for which we need the scalar and graviton wavefunctions. As in the gauge theory setting, the symmetries of the plane wave metric enable an explicit solution of the `free' equations of motion. These are again expressed in terms of the solution $\phi_k$ of the Hamilton-Jacobi equation~\cite{Ward:1987ws,Adamo:2017nia}:
\be\label{grHamJac}
\phi_{k}=k_{+}\,x^{+}+\frac{k_+}{2}\,\sigma_{ab}\,x^{a}x^{b}+k_{i}\,E^{i}_{a}\,x^{a}+\frac{1}{2\,k_+}\int^{x^{+}}\d s\,\left(m^2+k_i\,k_j\,\gamma^{ij}(s)\right)\,,
\ee
where $\{k_+, k_i\}$ are the free components of an on-shell momentum in $(d-2)$-dimensions. (We abuse notation, using $\phi_k$ for the solution to the Hamilton-Jacobi equations in both gauge theory and gravity; the distinction is always clear from the context.)

A massive scalar $\Phi$ is described by
\be\label{grScalar}
\Phi(x)=\Omega(x^-)\,\e^{\pm\im\,\phi_k}\,, \qquad (\nabla^2+m^2)\Phi=0\,,
\ee
where $\Omega(x^-):=|E|^{-1/2}$ and $\nabla^2=g^{\mu\nu}\,\nabla_{\mu}\nabla_{\nu}$ for $\nabla_{\mu}$ the Levi-Civita connection of the plane wave metric. Massless fields of higher spin can be obtained from \eqref{grScalar} using covariantly constant spin-raising operators~\cite{Mason:1989}. A photon $A_{\mu}$ in lightfront-Feynman gauge, $n\cdot A=0=\nabla\cdot A$, is given by
\be\label{grPhoton}
A_{\mu}=\cE_{\mu}\,\Omega\,\e^{\pm\im\,\phi_k}\,,
\ee
where $\cE_{\mu}$ is a dressed polarization vector and $m=0$ in the Hamilton-Jacobi function. The dressed polarization is given by the natural analogue of the projector \eqref{glpol}:
\be\label{grpol1}
\cE_{\mu}= \left(g_{\mu\nu}-\frac{K_{\mu}\,n_{\nu}+n_{\mu}\,K_{\nu}}{k_{+}}\right)\,\epsilon^{\nu}:=\mathbb{P}_{\mu\nu}\,\epsilon^{\nu}\,,
\ee
where the dressed momentum is given by $K_{\mu}:=\nabla_{\mu}\phi_{k}$ and obeys $K^2=0$. Note that unlike the gauge theory setting, $K_{\mu}$ -- and hence the projector $\mathbb{P}_{\mu\nu}$ -- is a function of the transverse coordinates $x^{\perp}$ as well as $x^-$. It is easy to check that $K\cdot\cE=0=n\cdot\cE$ everywhere.

For a spin-2 graviton, the dressed polarization picks up additional background dependence in the form of explicit `tail' terms. A graviton $h_{\mu\nu}$ in the plane wave space-time, subject to gauge conditions $n^{\mu}h_{\mu\nu}=0=h^{\mu}_{\mu}$ and $\nabla^{\mu}h_{\mu\nu}=0$, can be written
\be\label{grGraviton}
h_{\mu\nu}=\cE_{\mu\nu}\,\Omega\,\e^{\pm\im\,\phi_k}\,,
\ee
where the dressed polarization tensor is \emph{not} transverse: $K^{\mu}\cE_{\mu\nu}\neq0$. This is because $K^{\mu}\cE_{\mu\nu}=0$ is not equivalent to the covariant gauge-fixing condition $\nabla^{\mu}h_{\mu\nu}$ in the background. In terms of spin-1 projectors \eqref{grpol1}, the dressed polarization tensor is
\begin{align}\label{grpol2}
\begin{split}
\cE_{\mu\nu}&=\left(\mathbb{P}_{\mu\lambda}\,\mathbb{P}_{\nu\sigma}-\frac{\im}{k_+}\,n_{\mu}\,n_{\nu}\,\delta^{a}_{\lambda}\delta^{b}_{\sigma}\,\sigma_{ab}\right)\,\epsilon^{\lambda\sigma} =\mathbb{P}_{\mu\nu\lambda\sigma}\,\epsilon^{\lambda\sigma}\,,
\end{split}
\end{align}
where $\mathbb{P}_{\mu\nu\lambda\sigma}$ is the same spin-2 projector that we encountered before in \eqref{altTtilde*}. The un-dressed polarization $\epsilon_{\mu\nu}$ is assumed to be traceless with respect to the plane wave metric.

We are now ready to calculate the non-linear Compton amplitude. This is encoded by the minimal cubic coupling between a massive scalar and gravity:
\be\label{GRs1}
\kappa\,\int\d^{d}x\,\sqrt{|g|}\,h^{\mu\nu}\left(2\partial_{\mu}\Phi\,\partial_{\nu}\Phi-g_{\mu\nu}\left(\partial_{\rho}\Phi\,\partial^{\rho}\Phi-\frac{m^2}{2}\Phi^2\right)\right)\,,
\ee
where $g_{\mu\nu}$ is the background plane wave metric. Evaluating this integral on the wavefunctions \eqref{grScalar}, \eqref{grGraviton}, only the first term contributes due to the trace-free property of $h_{\mu\nu}$, leaving:
\be\label{GRs2}
-4\,\kappa\,\int \d^{d}x\,\Omega^{3}(x^-)\,\cE_{\mu\nu}\,P^{\mu}\,P^{\prime\,\nu}\,\exp\left[-\im\,\phi_{p}+\im\,\phi_{p'}+\im\,\phi_{k}\right]\,,
\ee
where $p_{\mu}$ and $p'_{\mu}$ are the incoming and outgoing scalar momenta, respectively ($p^2=m^2=p^{\prime\,2}$), $k_{\mu}$ is the emitted graviton momentum, and $\Omega=|E|^{-1/2}$. The dressed graviton polarization $\cE_{\mu\nu}$ is given by \eqref{grpol2}.

The integral in $\d x^{+}$ can be performed trivially to give a momentum conserving delta function $\delta(p'_{+}+k_{+}-p_{+})$, but it appears that the transverse integrals $\d^{d-2}x^{\perp}$ are obstructed since $\cE_{\mu\nu}$, $P_{\mu}$ and $P'_{\mu}$ all contain non-trivial $x^{\perp}$-dependence (in contrast to what occurs in the gauge theory background). However, an explicit calculation reveals that
\be\label{GRs3}
\cE_{\mu\nu}\,P^{\mu}\,P^{\prime\,\nu}=\left(\frac{E^{i}_{a}\,E^{j}_{b}}{k_{+}}\left(p_{+}\,k_{i}-k_{+}\,p_i\right)\left(p'_{+}\,k_{i}-k_{+}\,p'_i\right)-\im\,p_{+}\,p'_{+}\,\sigma_{ab}\right)\,\frac{\epsilon^{ab}}{k_+}\,.
\ee
In other words, the combination $\cE_{\mu\nu}\,P^{\mu}\,P^{\prime\,\nu}$ is a function of $x^-$ only. This enables the $\d^{d-2}x^{\perp}$ integrals to be performed, leaving
\be\label{GRnlc}
-4\,\kappa\,(2\pi)^{d-1}\,\delta^{+,\perp}(p'+k-p)\,\int\limits_{-\infty}^{+\infty}\frac{\d x^{-}}{\sqrt{|E|}}\,\cE_{\mu\nu}\,P^{\mu}\,P^{\prime\,\nu}(x^-)\,\e^{\im\,V[p,k]}\,,
\ee
where the gravitational Volkov exponent
\be\label{GRVolkov}
\begin{split}
V[p,k]&:=\int^{x^{\LCm}}\!\!\!\d s\, \frac{P_{\mu} K_{\nu}\,g^{\mu\nu}(s)}{(p-k)_{+}} \\
&=\frac{1}{(p-k)_+}\int^{x^-}\d s\,\gamma^{ij}(s)\left(\frac{p_{+}}{k_+}\,k_{i}k_{j}+\frac{k_+}{p_+}\,p_{i}p_{j}-p_i\,k_j\right)\,,
\end{split}
\ee
follows on the support of the momentum conserving delta functions.

\subsubsection*{Classical limit}

As before, the classical limit of the amplitude is taken by assuming that the emitted graviton momentum is negligible compared to that of the scalar ($k_{+} \ll p_\LCp$) and relaxing the constraints imposed by momentum conservation. In the gravitational Volkov exponent, the classical limit simply replaces $(p-k)_{\LCp}^{-1}$ with $p_{+}^{-1}$, but unlike the gluon emission amplitude, the tensor structure in \eqref{GRnlc} changes slightly in the classical limit. In particular, it follows that
\begin{multline}\label{GRclim1}
\cE_{\mu\nu}\,P^{\mu}\,P^{\prime\,\nu}=\left(\mathbb{P}_{\mu\lambda}\,\mathbb{P}_{\nu\sigma}\,P^{\mu}\,P^{\nu}-\im\,\frac{p_{+}\,(p-k)_{+}}{k_+}\,\sigma_{ab}\,\delta^{a}_{\lambda}\,\delta^{b}_{\sigma}\right)\epsilon^{\lambda\sigma}  \\
\xrightarrow{\mathrm{classical}}\mathbb{P}_{\mu\nu\lambda\sigma}\,P^{\mu}\,P^{\nu}\,\epsilon^{\lambda\sigma}\,,
\end{multline}
where all terms which are subleading in the classical limit have been dropped in passing to the second line.

Stripping the non-linear Compton amplitude of overall momentum conserving delta functions, and including factors for state normalisation, let
\be\label{GRclim2}
\mathcal{M}_{3}:=\frac{-4\,\kappa}{\sqrt{2k_{+}\,2p_{+}\,2(p-k)_{+}}}\int\limits_{-\infty}^{+\infty}\frac{\d x^{-}}{\sqrt{|E|}}\,\cE_{\mu\nu}\,P^{\mu}\,P^{\prime\,\nu}(x^-)\,\e^{\im\,V[p,k]}\,.
\ee
Then the classical limit obeys
\be\label{GRclim3}
\mathcal{M}_{3}\xrightarrow{\mathrm{classical}}\mathcal{H}(k)\,,
\ee
where $\mathcal{H}(k)$ is the radiation field at infinity \eqref{clGR3} generated by classical back reaction. Once again, we see that the classical limit of the non-linear Compton amplitude gives the correct classical result, including the appropriate state normalisation for the graviton radiation field.

\section{Double copy}
\label{SECT:DC}

In a trivial background, double copy relates perturbative scattering amplitudes in gauge and gravitational theories by the heuristic Gravity $=$ $(\text{Gauge})^2$ (cf., \cite{Bern:2019prr}). For 3-point amplitudes, this `formula' is literally true: if $A_3$ is a 3-point amplitude in some gauge theory (stripped of colour factors, coupling constants and momentum conserving delta functions) and $M_3$ is a 3-point amplitude in the appropriate\footnote{The `appropriate' gravitational theory is, generally speaking, the theory whose spectrum is the `square' of the gauge theory spectrum. For example, the appropriate gravitational theory for pure Yang-Mills theory is the NS-NS sector of supergravity, containing a metric, dilaton and $B$-field.} gravitational theory (stripped of coupling constants and momentum conserving delta functions), then the two are related by $A_{3}^2=M_3$. 

For amplitudes on a plane wave background, the situation is complicated by the fact that there is only $(d-1)$-dimensional momentum conservation, so it is not immediately clear what to use as the `stripped' amplitudes. Any 3-point amplitude in a plane wave background has the general structure
\be\label{3pstruct}
g\,(2\,\pi)^{d-1}\,\delta^{+,\perp}(p'+k-p)\,\int\limits_{-\infty}^{+\infty}\d \mu\,\mathcal{I}_{3}(x^{-})\,\e^{\varphi(x^-)}\,,
\ee
where $g$ is the appropriate coupling constant; $\mathcal{I}_{3}$ is the `tree-level integrand' comprising the interesting kinematical content of the amplitude; $\d\mu$ is a theory-dependent measure (which may include colour factors) in the lightfront time $x^-$; and $\varphi$ is the theory-dependent Volkov exponent. For instance, on a gauge theory plane wave background,
\be\label{gtmeas}
\d\mu|_{\mathrm{YM}}=\mathrm{f}^{\mathsf{abc}}\,\d x^{-}\,, \qquad \varphi|_{\mathrm{YM}}=\im\,\mathcal{V}[p,k]\,,
\ee
while on a gravitational plane wave background
\be\label{grmeas}
\d\mu|_{\mathrm{GR}}=\frac{\d x^{-}}{\sqrt{|E|}}\,, \qquad \varphi|_{\mathrm{GR}}=\im\,V[p,k]\,.
\ee
The ingredients $\d\mu$ and $\varphi$ in \eqref{3pstruct} are universal: given the background and the asymptotic kinematics, they can be written down without any further thought or calculation.

Thus, the objects which play the role of stripped amplitudes and should be related by double copy on a plane wave background are the tree-level integrands, $\mathcal{I}_{3}$. In~\cite{Adamo:2017nia}, a proposal for this double copy was put forward, consisting of three steps. Let $\mathcal{A}_{3}$ be the (given) tree-level integrand of a 3-point gauge theory amplitude on a Yang-Mills plane wave background, and $\mathcal{M}_3$ be the (desired) tree-level integrand of a 3-point amplitude in the appropriate gravitational theory on a plane wave space-time. The prescription of~\cite{Adamo:2017nia} is:
\begin{enumerate}
 \item Define $\widetilde{\mathcal{A}}_{3}$ by reversing the sign all charges with respect to the background for each particle (i.e., $e_{A}\rightarrow-e_{A}$ for $A=p,k,p'$), and take
 \begin{equation*}
 |\mathcal{A}_{3}|^{2}:=\mathcal{A}_{3}\,\widetilde{\mathcal{A}}_{3}\,.
 \end{equation*}
 
 \item In $|\mathcal{A}_3|^2$, replace scalar products of Minkowski kinematics with curved space-time kinematics, according to the rule
 \begin{equation*}
  \epsilon_{A}\cdot k_{B}\rightarrow \epsilon^{\mu}_{A}\,\mathbb{P}_{\mu\nu}\,K_{B}^{\nu}\,,
 \end{equation*}
 where $\mathbb{P}_{\mu\nu}$ is the spin-1 projector \eqref{grpol1*} and $K^{\mu}_{B}$ is the dressed momentum of particle $B$ in the plane wave space-time. For terms quadratic in a polarization vector, replace $\epsilon_{\mu}\,\epsilon_{\nu}\rightarrow \epsilon_{\mu\nu}$. 
 
 \item Replace all occurences of the gauge theory plane wave $a_{\perp}(x^-)$ and colour charges $e_A$ by
 \begin{equation*}
 e_{A}\,e_{B}\,a_{a}(x^-)\,a_{b}(x^-)\rightarrow\left\{\begin{array}{rl}
                                                                            -\im\,\sigma_{ab}(x^-)\,\mathfrak{s}(A)\,k_{A\,+} & \mbox{if } A=B \\
                                                                            \im\,\sigma_{ab}(x^-)\left(\mathfrak{s}(A)\,k_{A\,+} +\mathfrak{s}(B)\,k_{B\,+}\right) & \mbox{otherwise}
                                                                            \end{array}\right.\,
 \end{equation*}
 where $\mathfrak{s}(A)$ is $-1$ if particle $A$ is incoming and $+1$ if outgoing.
\end{enumerate}

Abbreviating steps (2-3) of the above algorithm by the replacement map $\rho$, this double copy prescription can be succinctly stated as:
\be\label{DCprescrip}
\mathcal{M}_{3}=\rho\left(|\mathcal{A}_{3}|^2\right)\,.
\ee
In~\cite{Adamo:2017nia}, this was tested for the three-point gluon and graviton amplitudes on plane wave backgrounds, and we can now apply it to the non-linear Compton amplitudes.

In our case, $\mathcal{A}_{3}$ is the tree-level integrand of the gluon emission amplitude for a massive colour-charged scalar crossing the Yang-Mills plane wave background\footnote{Here, we have dropped an overall numerical factor of $2\,\im$ which trivally squares to the overall numerical factor $-4$ in the gravity amplitude}, recall (\ref{YMs3}):
\be\label{DC1}
\mathcal{A}_{3} \equiv\mathcal{E} \cdot P = \epsilon\cdot p +\frac{\epsilon\cdot a}{k_+}\left(e_{k}\,p_{+}-e_{p}\,k_{+}\right)\,.
\ee
Then following step (1),
\be\label{DC2}
|\mathcal{A}_3|^2=(\epsilon\cdot p)^{2}-\frac{(\epsilon\cdot a)^2}{k_{+}^2}\left(e_{p}^2\,k_{+}^2+e_{k}^{2}\,p_{+}-2\,e_{k}e_{p}\,k_{+}p_{+}\right)\,,
\ee
and step (2) gives
\be\label{DC3}
\epsilon^{\lambda\sigma}\,\mathbb{P}_{\mu\sigma}\,\mathbb{P}_{\nu\lambda}\,P^{\mu}\,P^{\nu}-\frac{\epsilon^{ab}\,a_{a}\,a_{b}}{k_{+}^2}\left(e_{p}^2\,k_{+}^2+e_{k}^{2}\,p_{+}-2\,e_{k}e_{p}\,k_{+}p_{+}\right)\,.
\ee
Replacing the gauge field background and colour charges via step (3) results in
\be\label{DC4}
\epsilon^{\lambda\sigma}\,\mathbb{P}_{\mu\sigma}\,\mathbb{P}_{\nu\lambda}\,P^{\mu}\,P^{\nu}-\frac{\im}{k_+}\,\epsilon^{ab}\sigma_{ab}\,p_{+}\,(p-k)_+ =\cE_{\mu\nu}\,P^{\mu}\,P^{\prime\,\nu}\,.
\ee
This is precisely the tree-level integrand of \eqref{GRnlc}, the non-linear Compton amplitude for graviton emission. So the double copy prescription also works for 3-point amplitudes with two massive scalar legs in plane wave backgrounds. 

\medskip

Since the back-reaction computed in Sections~\ref{SECT:YM} and~\ref{SECT:GR} is captured by the classical limit of the non-linear Compton amplitudes, double copy at the level of the amplitudes implies double copy at the level of the classical radiation field. Nevertheless, it is instructive to note that a set of replacement rules can be given at the classical level which generalize those previously derived for the perturbative double copy with no background field~\cite{Goldberger:2016iau,Luna:2016hge,Goldberger:2017vcg}, as follows. In the expression \eqref{Classical-result} for the classical gluon radiation field at infinity $\mathcal{A}^{\sfa}$, the replacements
\be\label{DC5}
c^{\sfa}_{\mathrm{in}} \rightarrow P^{\prime\nu}\,, \qquad \mathbb{P}_{\mu\nu} \rightarrow \mathbb{P}_{\mu\nu\lambda\sigma}\,,
\ee
along with $P\cdot K\rightarrow g^{\mu\nu}P_{\mu}\,K_{\nu}$ in the exponent convert $\mathcal{A}^{\sfa}\rightarrow \mathcal{H}$, where $\mathcal{H}$ is the classical graviton radiation field at infinity given by \eqref{clGR3}. The rules \eqref{DC5} are simply background-dressed versions of those governing the perturbative double copy in the absence of strong background fields.

\section{Four-dimensions and spinor helicity}
\label{SECT:SHF}

In four space-time dimensions the Lorentz group is locally isomorphic to SL$(2,\mathbb{C})$, a fact with important implications for on-shell kinematics. One of these is the existence of the \emph{spinor helicity formalism}, which gives an unconstrained method to generate on-shell kinematics in terms of spinor variables. The spinor helicity formalism for massless fields in a trivial background is now a widely used tool in the scattering amplitudes community (cf., \cite{Dixon:2013uaa,Cheung:2017pzi}). In~\cite{Adamo:2019zmk}, we showed that the spinor helicity formalism extends naturally to massless fields in plane wave backgrounds, and this plays a crucial role in new all-multiplicity formulae for gluon and graviton scattering in chiral plane waves~\cite{Adamo:2020syc}.

Here, we observe that the spinor helicity formalism for \emph{massive} particles also extends naturally to describe on-shell kinematics in a plane wave background. While the `massive' spinor helicity formalism on a trivial background has existed in various guises for decades (cf., \cite{Penrose:1972ia,Perjes:1974ra,Tod:1976sk,Kleiss:1986qc,Conde:2016vxs,Conde:2016izb,Weinzierl:2016bus}), we follow the notation and conventions of~\cite{Arkani-Hamed:2017jhn}. After setting out the formalism for both gauge theory and gravity, we then translate the non-linear Compton amplitudes from Section~\ref{NLComp} into this `dressed' spinor helicity formalism. 


\subsection{The formalism}
\label{SSECT:form}

Restricting to four space-time dimensions, the Minkowski metric can be written as
\be\label{4dMink}
\d s^{2}=2\left(\d x^{+}\,\d x^{-}-\d z\,\d\bar{z}\right)\,,
\ee
where the transverse coordinates are identified with the complex plane. The complexified Lorentz group obeys SO$(4,\mathbb{C})\cong$ SL$(2,\mathbb{C})\times$SL$(2,\mathbb{C})$ locally. In practical terms, this means that any vector index can be traded for a pair of SL$(2,\mathbb{C})$ spinor indices (one of each chirality), and this is operationalized by contraction with the Pauli matrices: $v^{\alpha\dot\alpha}=\sigma_{\mu}^{\alpha\dot\alpha}\,v^{\mu}$. In spinor variables, the coordinates on Minkowski space can be packaged into
\be\label{lfcs}
x^{\alpha\dot\alpha}=\left(\begin{array}{cc}
										x^{+} & \bar{z} \\
										z & x^{-}
										\end{array}\right)\,,
\ee 
and the Minkowski metric is $\d s^2=\d x_{\alpha\dot\alpha}\,\d x^{\alpha\dot\alpha}$, where spinor indices are raised and lowered using the Levi-Civita symbols $\epsilon_{\alpha\beta}$, $\epsilon^{\alpha\beta}$, etc. Our conventions are $a^{\alpha}=\epsilon^{\alpha\beta}\,a_{\beta}$, $a_{\alpha}=a^{\beta}\,\epsilon_{\beta\alpha}$, and similarly for dotted spinor indices.

It is a simple fact of linear algebra that if $k^{\alpha\dot\alpha}$ is null ($k^2=0$), then it can be decomposed into spinors as\footnote{In this section, we allow for all momenta to be complex, so momentum spinors (massless or massive) of opposite chirality are not assumed to be related by complex conjugation.} $k_{\alpha\dot\alpha}^{\mathrm{null}}=\lambda_{\alpha}\tilde{\lambda}_{\dot\alpha}$. Similarly, any non-null vector $k^{\alpha\dot\alpha}$, with $k^2=m^2$, can be written as a sum of two null vectors:
\be\label{sh1}
\begin{split}
k_{\alpha\dot\alpha}&=\lambda^{1}_{\alpha}\,\tilde{\lambda}_{\dot\alpha\,1}+\lambda^{2}_{\alpha}\,\tilde{\lambda}_{\dot\alpha\,2} \\
 &:=\lambda^{\mathrm{a}}_{\alpha}\,\tilde{\lambda}_{\dot\alpha\,\mathrm{a}}=\lambda^{\mathrm{a}}_{\alpha}\,\tilde{\lambda}^{\mathrm{b}}_{\dot\alpha}\,\epsilon_{\mathrm{ba}}\,,
\end{split}
\ee
where the indices $\mathrm{a},\mathrm{b},\ldots=1,2$ are identified with the little group SO$(3)\cong$ SU$(2)$ for massive particles in four-dimensions. These little group spinors obey
\be\label{sh2}
k_{\alpha\dot\alpha}\,\tilde{\lambda}^{\dot\alpha\,\mathrm{a}}=m\,\lambda^{\mathrm{a}}_{\alpha}\,, \qquad k_{\alpha\dot\alpha}\,\lambda^{\alpha\,\mathrm{a}}=-m\,\tilde{\lambda}^{\mathrm{a}}_{\dot\alpha}\,,
\ee
which are equivalent to the Dirac equation for a momentum eigenstate. 

The decomposition of a momentum vector into spinors is just a fact of linear algebra, so it must hold for the dressed momenta of massless and massive particles in a plane wave background, since the dressed momenta remain on-shell. For a gauge theory plane wave background in 4-dimensions, the two Cartan-valued background degrees of freedom are encoded in
\be\label{gsback}
a_{\alpha\dot\alpha}(x^-)=\left(\begin{array}{cc}
														0 & \tilde{a}(x^-) \\
														a(x^-) & 0
										\end{array}\right)\,
\ee
where the background is allowed to be complex; real-valued backgrounds are recovered by setting $\tilde{a}=\bar{a}$. By virtue of $n^2=0$, the vector $n^{\alpha\dot\alpha}$ associated with plane wave background has a spinor decomposition
\be\label{nspinor}
n^{\alpha\dot\alpha}=\iota^{\alpha}\,\tilde{\iota}^{\dot\alpha}\,, \qquad \iota^{\alpha}=\left(\begin{array}{c} 1 \\ 0 \end{array}\right)=\tilde{\iota}^{\dot\alpha}\,.
\ee
With these ingredients, the dressed momentum of a (massive or massless) particle in the plane wave background reads
\be\label{dsh1}
K_{\alpha\dot\alpha}(x^-)=k_{\alpha\dot\alpha}-e\,a_{\alpha\dot\alpha} +\frac{\iota_{\alpha}\,\tilde{\iota}_{\dot\alpha}}{2\,k_+}\,(2\,e\,a\cdot k-e^2\,a^2)\,,
\ee
where $e$ is the charge of the particle with respect to the background.

In the massless case ($k^2=0=K^2$), it was shown~\cite{Adamo:2019zmk} that 
\be\label{dshmassless}
K^{\mathrm{null}}_{\alpha\dot\alpha}(x^-)=\Lambda_{\alpha}(x^-)\,\tilde{\Lambda}_{\dot\alpha}(x^-)\,,
\ee
with the dressed spinors given by 
\be\label{dspinmassless}
\Lambda_{\alpha}=\lambda_{\alpha}-\frac{e\,a(x^-)}{[\tilde{\iota}\,\tilde{\lambda}]}\,\iota_{\alpha}\,, \qquad \tilde{\Lambda}_{\dot\alpha}=\tilde{\lambda}_{\dot\alpha}-\frac{e\,\tilde{a}(x^-)}{\la\iota\,\lambda\ra}\,\tilde{\iota}_{\dot\alpha}\,.
\ee
Here, the spinors $\lambda_{\alpha}$, $\tilde{\lambda}_{\dot\alpha}$ are those of the (constant) null momentum before entering the background: $k_{\alpha\dot\alpha}^{\mathrm{null}}=\lambda_{\alpha}\tilde{\lambda}_{\dot\alpha}$. We adopt the usual notation for SL$(2,\mathbb{C})$-invariant contractions of constant spinors
\be\label{spinconv}
\la a\,b\ra:=a^{\alpha}\,b_{\alpha}=\epsilon^{\alpha\beta}\,a_{\beta}\,b_{\alpha}\,, \qquad [\tilde{a}\,\tilde{b}]:=\tilde{a}^{\dot\alpha}\,\tilde{b}_{\dot\alpha}=\epsilon^{\dot\alpha\dot\beta}\,\tilde{a}_{\dot\beta}\,\tilde{b}_{\dot\alpha}\,,
\ee
and define a `double-bracket' notation to distinguish contractions between \emph{dressed} spinors:
\be\label{dspinconv}
\la\!\la A\,B\ra\!\ra:=A^{\alpha}(x)\,B_{\alpha}(x)\,, \qquad [\![\tilde{A}\,\tilde{B}]\!]:=\tilde{A}^{\dot\alpha}(x)\,\tilde{B}_{\dot\alpha}(x)\,.
\ee
This will prove convenient to distinguish between constant and dressed spinor contractions in subsequent calculations.

Now, for a \emph{massive} dressed momentum \eqref{dsh1} holds with $k^2=m^2$, and linear algebra implies that
\be\label{dsh2}
K_{\alpha\dot\alpha}(x^-)=\Lambda^{\mathrm{a}}_{\alpha}(x^-)\,\tilde{\Lambda}_{\dot\alpha\,\mathrm{a}}(x^-)\,.
\ee
Combined with \eqref{dsh1}, this enables us to fix the precise form of massive dressed momentum spinors:
\be\label{dspinors}
 \Lambda^{\mathrm{a}}_{\alpha}=\lambda^{\mathrm{a}}_{\alpha}-\frac{e\,a}{k_+}\,\iota_{\alpha}\,\iota^{\beta}\,\lambda_{\beta}^{\mathrm{a}}\,, \qquad \tilde{\Lambda}^{\mathrm{a}}_{\dot\alpha}=\tilde{\lambda}^{\mathrm{a}}_{\dot\alpha}-\frac{e\,\tilde{a}}{k_+}\,\tilde{\iota}_{\dot\alpha}\,\tilde{\iota}^{\dot\beta}\,\tilde{\lambda}_{\dot\beta}^{\mathrm{a}}\,,
\ee
with $k_{\alpha\dot\alpha}=\lambda^{\mathrm{a}}_{\alpha}\tilde{\lambda}_{\dot\alpha\,\mathrm{a}}$. It is straightforward to check that $K^2=m^2$ and 
\be\label{dsh3}
K_{\alpha\dot\alpha}\,\tilde{\Lambda}^{\dot\alpha\,\mathrm{a}}=m\,\Lambda^{\mathrm{a}}_{\alpha}\,, \qquad K_{\alpha\dot\alpha}\,\Lambda^{\alpha\,\mathrm{a}}=-m\,\tilde{\Lambda}^{\mathrm{a}}_{\dot\alpha}\,.
\ee
Note that in both the massless \eqref{dspinmassless} and massive \eqref{dspinors} settings, there is manifest chirality in the dressing of the spinors: $\Lambda_{\alpha}$ and $\Lambda^{\mathrm{a}}_{\alpha}$ depend only on $a(x^-)$, while $\tilde{\Lambda}_{\dot\alpha}$ and $\tilde{\Lambda}^{\mathrm{a}}_{\dot\alpha}$ depend only on $\tilde{a}(x^-)$.

\medskip

A similar -- albeit slightly more complicated -- story holds for on-shell kinematics in a gravitational plane wave space-time. In four-dimensions, the (complexified) plane wave metric takes the form:
\be\label{4dmet}
\d s^{2}=2\left(\d x^{+}\,\d x^{-}-\d z\,\d\tilde{z}\right)-\left[z^{2}\,\ddot{f}(x^-)+\tilde{z}^2\,\ddot{\tilde{f}}(x^-)\right]\,(\d x^-)^2\,,
\ee
where $f(x^-)$, $\tilde{f}(x^-)$ are the two functional degrees of freedom, and the Lorentzian-real space-time is recovered by imposing the reality conditions $\tilde{z}=\bar{z}$ and $\tilde{f}=\bar{f}$. The Ricci curvature of this metric vanishes (as required by the vacuum Einstein equations), and the Weyl curvature can be decomposed into its self-dual and anti-self-dual parts, written in SL$(2,\mathbb{C})$ spinors as:
\be\label{Weyl}
\widetilde{\Psi}_{\dot\alpha\dot\beta\dot\gamma\dot\delta}=\tilde{\iota}_{\dot\alpha}\tilde{\iota}_{\dot\beta}\tilde{\iota}_{\dot\gamma}\tilde{\iota}_{\dot\delta}\,\ddot{\tilde{f}}\,, \quad \Psi_{\alpha\beta\gamma\delta}=\iota_{\alpha}\iota_{\beta}\iota_{\gamma}\iota_{\delta}\,\ddot{f}\,.
\ee
So $\tilde{f}$ and $f$ control the self-dual and anti-self-dual curvature of the plane wave metric, respectively.

The frames $E^{i}_{a}$ and deformation tensor $\sigma_{ab}$ which played an important role in general dimension are captured by the transverse frame
\be\label{frame}
E=\d z+f(x^-)\,\d\tilde{z}\,, \qquad \tilde{E}=\d \tilde{z}+\tilde{f}(x^-)\,\d z\,,
\ee
and shear components
\be\label{shear}
\sigma(x^-)=\dot{f}(x^-)\,, \qquad \tilde{\sigma}(x^-)=\dot{\tilde{f}}(x^-)\,.
\ee
The curved metric \eqref{4dmet} can be expressed as $\d s^2=\epsilon_{\alpha\beta}\epsilon_{\dot\alpha\dot\beta}\,e^{\alpha\dot\alpha}e^{\beta\dot\beta}$ in terms of the tetrad
\be\label{tetrad}
e^{\alpha\dot\alpha}=\d x^{\alpha\dot\alpha}-\frac{\iota^{\alpha}\,\tilde{\iota}^{\dot\alpha}}{2}\left(z^{2}\,\ddot{f}(x^-)+\tilde{z}^2\,\ddot{\tilde{f}}(x^-)\right)\,\d x^{-}\,.
\ee
The dual tetrad, 
\be\label{dtetrad}
e_{\alpha\dot\alpha}=\partial_{\alpha\dot\alpha}+\frac{\iota_{\alpha}\,\tilde{\iota}_{\dot\alpha}}{2}\left(z^{2}\,\ddot{f}(x^-)+\tilde{z}^2\,\ddot{\tilde{f}}(x^-)\right)\,\partial_{+}\,,
\ee
obeys $e_{\alpha\dot\alpha}\lrcorner e^{\beta\dot\beta}=\delta_{\alpha}^{\beta}\delta^{\dot\beta}_{\dot\alpha}$, and the spinor components of any vector or 1-form are defined by its expansion in these bases.

Consequently, if $k^{\mathrm{null}}_{\alpha\dot\alpha}=\lambda_{\alpha}\tilde{\lambda}_{\dot\alpha}$ is a massless momentum prior to entering the curved region of the sandwich plane wave, then the dressed null momentum is
\be\label{grmassless}
K^{\mathrm{null}}_{\alpha\dot\alpha}(x^-)=\Lambda_{\alpha}(x^-)\,\tilde{\Lambda}_{\dot\alpha}(x^-)\,,
\ee
with the dressed spinors given by 
\be\label{grspinmassless}
\begin{split}
\Lambda_{\alpha}&=\lambda_{\alpha}-\frac{\la\iota\,\lambda\ra}{[\tilde{\iota}\,\tilde{\lambda}]}\,\iota_{\alpha}\left([\tilde{o}\,\tilde{\lambda}]\,f+[\tilde{\iota}\,\tilde{\lambda}]\,z\,\dot{f}\right)\, \\
\tilde{\Lambda}_{\dot\alpha}&=\tilde{\lambda}_{\dot\alpha}-\frac{[\tilde{\iota}\,\tilde{\lambda}]}{\la\iota\,\lambda\ra}\,\left(\la o\,\lambda\ra\,\tilde{f}+\la\iota\,\lambda\ra\,\tilde{z}\,\dot{\tilde{f}}\right)\,.
\end{split}
\ee
Here, we have introduced the constant spinors $o^{\alpha}$, $\tilde{o}^{\dot\alpha}$ defined by $\la \iota\,o\ra=1=[\tilde{\iota}\,\tilde{o}]$. For a massive initial momentum $k_{\alpha\dot\alpha}=\lambda^{\mathrm{a}}_{\alpha}\tilde{\lambda}_{\dot\alpha\,\mathrm{a}}$, the background dressed momentum becomes
\be\label{grds1}
K_{\alpha\dot\alpha}=\Lambda^{\mathrm{a}}_{\alpha}\,\tilde{\Lambda}_{\dot\alpha\,\mathrm{a}}\,, 
\ee
where the dressed spinors are:
\be\label{grds2}
\begin{split}
\Lambda^{\mathrm{a}}_{\alpha}&=\lambda^{\mathrm{a}}_{\alpha}-\frac{\iota_{\alpha}}{k_+}\,\iota^{\beta}\,\lambda^{\mathrm{a}}_{\beta}\left(k_{+}\,z\,\dot{f}-\tilde{k}\,f\right)\,, \\
\tilde{\Lambda}^{\mathrm{a}}_{\dot\alpha}&=\tilde{\lambda}^{\mathrm{a}}_{\dot\alpha}-\frac{\tilde{\iota}_{\dot\alpha}}{k_+}\,\tilde{\iota}^{\dot\beta}\,\tilde{\lambda}^{\mathrm{a}}_{\dot\beta}\left(k_{+}\,\tilde{z}\,\dot{\tilde{f}}-k\,\tilde{f}\right)\,.
\end{split}
\ee
Here, we abbreviate the transverse intitial momentum components by
\be\label{ctransverse}
k:=o^{\alpha}\,\lambda^{\mathrm{a}}_{\alpha}\tilde{\lambda}^{\dot\alpha}_{\mathrm{a}}\,\tilde{\iota}_{\dot\alpha}\,, \qquad \tilde{k}:=\iota^{\alpha}\,\lambda^{\mathrm{a}}_{\alpha}\tilde{\lambda}^{\dot\alpha}_{\mathrm{a}}\,\tilde{o}_{\dot\alpha}\,,
\ee
since there is no notion of little group `weight' in the massive case, and nothing is gained by expressing these components in terms of spinor contractions. The dressed momentum spinors obey the relations
\be\label{grdirac}
K_{\alpha\dot\alpha}\,\tilde{\Lambda}^{\dot\alpha\,\mathrm{a}}=m\,\Lambda^{\mathrm{a}}_{\alpha}\,, \qquad K_{\alpha\dot\alpha}\,\Lambda^{\alpha\,\mathrm{a}}=-m\,\tilde{\Lambda}^{\mathrm{a}}_{\dot\alpha}\,,
\ee
which are equivalent to the Dirac equations on the curved space-time manifold.


\subsection{4d non-linear Compton amplitudes}

It is now straightforward to translate the non-linear Compton amplitudes \eqref{YMs3} and \eqref{GRnlc} into the dressed spinor helicity formalism. To do this, it is convenient to change notation slightly by labeling particle momenta with numbers: the incoming massive/charged scalar has momentum $k_1$, the outgoing massive/charged scalar has momentum $k_2$ and the emitted gauge boson has momentum $k_3$. In this notation, the amplitude for gluon emission becomes
\be\label{4dym1}
2\im\,g\,\int \d x^{-}\,\cE_{3}\cdot K_{1}(x^-)\,\e^{\im\,\mathcal{V}[1,3]}\,,
\ee
where factors associated to the colour and momentum conservation in the $x^+$, $z$ and $\tilde{z}$-directions have been dropped. Suppose that the emitted gluon has positive helicity; its dressed polarization vector is~\cite{Adamo:2019zmk}
\be\label{posglpol}
\cE^{(+)}_{3\,\alpha\dot\alpha}=\frac{\iota_{\alpha}\,\tilde{\Lambda}_{3\,\dot\alpha}}{\la\iota\,3\ra}\,.
\ee
Then the tensor structure of \eqref{4dym1} becomes
\be\label{4dym3}
\cE^{(+)}_{3}\cdot K_{1}(x^-)=\frac{\la\iota|K_1|3]\!]}{\la\iota\,3\ra}(x^-)=m\,\mathcal{X}(x^-)\,,
\ee
where $\mathcal{X}(x^-)$ is a dressed version of the `$x$ factor' introduced in~\cite{Arkani-Hamed:2017jhn} to describe minimal cubic couplings:
\be\label{xfactor}
\mathcal{X}(x^-):=\frac{\la\iota|K_1|3]\!]}{m\,\la\iota\,3\ra}(x^-)\,.
\ee
This is the obvious background-dressed version of the 3-point gluon emission amplitude on a trivial background given in~\cite{Arkani-Hamed:2017jhn}.


\medskip

Likewise, the non-linear Compton amplitude for graviton emission \eqref{GRnlc} is
\be\label{4dgr1}
-4\kappa\,\int \frac{\d x^{-}}{\sqrt{1-|f|^2}}\,\cE_{3\,\mu\nu}\,K_{1}^{\mu} K_{2}^{\nu}(x^-)\,\e^{\im\,V[1,3]}\,,
\ee
having dropped factors related to momentum conservation. The dressed polarization for a positive helicity emitted graviton is~\cite{Adamo:2020syc}
\be\label{posgrpol}
\cE^{(+)}_{\alpha\dot\alpha\beta\dot\beta}=\frac{\iota_{\alpha}\,\iota_{\beta}}{\la\iota\,3\ra^2}\left(\tilde{\Lambda}_{3\,\dot\alpha}\,\tilde{\Lambda}_{3\,\dot\beta}+\im\,\dot{\tilde{f}}\,\tilde{\iota}_{\dot\alpha}\,\tilde{\iota}_{\dot\beta}\right)\,,
\ee
for which the tensor structure of \eqref{4dgr1} becomes
\be\label{4dgr3}
\begin{split}
\cE^{(+)}_{3\,\mu\nu}\,K_{1}^{\mu}\,K_{2}^{\nu}&=\frac{\la\iota|K_1|3]\!]^2}{\la\iota\,3\ra^2}+\im\,\dot{\tilde{f}}\,\frac{k_{1\,+}\,(k_1-k_3)_{+}}{k_{3\,+}} \\
 & = m^2\,\mathcal{X}^{2}(x^-)+\im\,\dot{\tilde{f}}\,\frac{k_{1\,+}\,(k_1-k_3)_{+}}{k_{3\,+}}\,,
\end{split}
\ee
where the first equality follows on the support of momentum conservation and $\mathcal{X}$ is simply \eqref{xfactor} evaluated on the appropriate gravitationally-dressed spinors. 

Unlike the gluon emission amplitude \eqref{4dym3}, this differs from the answer obtained by naively `dressing' the flat background result~\cite{Arkani-Hamed:2017jhn}. Indeed, the naive dressing would only produce the first term in \eqref{4dgr3}; the second term arises from explicit tail effects in the gravitational setting.

\medskip

It is instructive to see how the double copy prescription for 3-point amplitudes works in terms of the spinor helicity variables. The tree-level integrand for gluon emission is
\be\label{4ddc1}
\mathcal{A}_{3}=m\,\mathcal{X}=m\,x+\frac{\tilde{a}}{k_{3\,+}}\,\left(e_{3}\,k_{1\,+}- e_{1}\,k_{3\,+}\right)\,,
\ee
where $x$ is the un-dressed $x$ factor of~\cite{Arkani-Hamed:2017jhn}. Following the prescription of Section~\ref{SECT:DC} we form
\be\label{4ddc2}
|\mathcal{A}_3|^2=m^2\,x^2-\frac{\tilde{a}^2}{k_{3\,+}^{2}}\,\left(e_{3}^{2}\,k_{1\,+}^{2}+e_{1}^{2}\,k_{3\,+}^{2}-2e_{1}e_{3}\,k_{1\,+}\,k_{3\,+}\right)\,,
\ee
and then apply the kinematics and background replacement map to find
\be\label{4ddc3}
\rho\left(|\mathcal{A}_3|^2\right)=m^2\,\mathcal{X}^2+\im\,\frac{\tilde{\sigma}}{k_{3\,+}}\,k_{1\,+}\,(k_1-k_3)_+\,,
\ee
where $\mathcal{X}$ is now the gravitationally-dressed $x$ factor. Using $\tilde{\sigma}=\dot{\tilde{f}}$ from \eqref{shear}, we recover the tree-level integrand for graviton emission, as desired.

\section{Future directions}
\label{SECT:CONCS}

In this paper, we have computed, both classically and in quantum field theory, leading gluon and graviton emission from charged/massive scalar particles in plane wave backgrounds, and shown that the results are related by double copy. In the special case of four space-time dimensions, a background dressed version of the spinor helicity formalism was developed and used to simplify the form of the non-linear Compton scattering amplitudes. 

There are many interesting directions for future research suggested by these results. The first of these is the inclusion of spin for the charged/massive probe particle. From the classical perspective, this can be done using the formalism of~\cite{Goldberger:2017ogt,Maybee:2019jus} with the inclusion of a strong plane wave background, while background-dressed QFT wavefunctions for spin-1/2 fields were developed in~\cite{Adamo:2019zmk}. In four-dimensions, the dressed spinor-helicity formalism could also enable the calculation of back-reaction from a Kerr black hole traversing a plane wave background. In a trivial background, the spin-dependence of 3-point amplitudes between two massive spin-$s$ particles and an emitted graviton exponentiates~\cite{Guevara:2018wpp,Chung:2018kqs}, and in the $s\rightarrow\infty$ limit reproduces all of the spin-induced multipoles of the Kerr black hole~\cite{Vines:2017hyw,Guevara:2019fsj,Bern:2020buy}. The presence of tail terms in \eqref{4dgr3} -- which surely persists at higher spin -- will introduce new subtleties in the presence of background fields.

It would also be interesting to investigate loop corrections to back-reaction, as well as considering higher-multiplicity scattering in a plane wave background. It is well-known from corresponding QED calculations that the complexity of higher-multiplicity amplitudes in plane wave backgrounds increases dramatically with the number of external legs. Although new approximations for the calculation of such amplitudes continue to be developed~\cite{Dinu:2019pau}, to date the only exact expressions available are for a few four-point amplitudes (e.g.~\cite{Dinu:2019pau,Bragin:2020akq} and references therein). However, all of these prior investigations have relied on traditional space-time Feynman diagrams. In the purely massless sector, formulae for tree-level MHV scattering at \emph{arbitrary} multiplicity on a chiral (self-dual) plane wave background were recently discovered, using integrability methods based on twistor theory~\cite{Adamo:2020syc}. Perhaps similar all-multiplicity statements are possible with massive external legs in such chiral backgrounds.

Building on~\cite{Adamo:2019zmk}, the discussion in Section~\ref{SSECT:form} sets out the ingredients of the spinor helicity formalism in any four-dimensional gauge or gravitational plane wave background. Separately from the other topics considered here, this raises the possibility of undertaking `standard' computations in strong-field QED using the spinor helicity formalism. Given the widespread utility of this formalism in the study of perturbative QFT on trivial backgrounds, it seems plausible to hope that it could have a similar positive impact in strong backgrounds.


\acknowledgments

We thank T. Heinzl, A. MacLeod \& D. O'Connell for helpful comments and conversations. TA is supported by a Royal Society University Research Fellowship. AI is supported by the EPSRC, grant EP/S010319/1.

\bibliography{DoubleCopyComptonBib}
\bibliographystyle{JHEP}
 
\end{document}